\documentclass{aa}  
\usepackage{natbib}
\usepackage{graphicx}
\usepackage{txfonts}
\usepackage{ulem}
\usepackage{threeparttable}
\usepackage{mathrsfs}
\usepackage{amsmath}

\begin{document}

   \title{The LOFAR and JVLA view of the distant steep spectrum radio halo in MACS J1149.5+2223}

   \author{L. Bruno 
          \inst{1,2},
          K. Rajpurohit
          \inst{2,1},
          G. Brunetti
         \inst{1},
         F. Gastaldello
         \inst{3},
         A. Botteon
          \inst{4},
           A. Ignesti
          \inst{5,2},
          A. Bonafede
         \inst{2,1},
          D. Dallacasa
          \inst{2,1},
          R. Cassano
         \inst{1},
         R. J. van Weeren
         \inst{4},
         V. Cuciti
         \inst{6},
         G. Di Gennaro
         \inst{4},
         T. Shimwell
         \inst{7,4},
         \and
         M. Br\"uggen
         \inst{6}
          }

   \institute{Istituto Nazionale di Astrofisica (INAF) - Istituto di Radioastronomia (IRA), via Gobetti 101, 40129 Bologna, Italy
   \and
   Dipartimento di Fisica e Astronomia (DIFA), Universit\`a di Bologna, via Gobetti 93/2, 40129 Bologna, Italy
             \and
             INAF - IASF Milano, Via A. Corti 12, I-20133, Milano, Italy
             \and
             Leiden Observatory, Leiden University, PO Box 9513, 2300 RA Leiden, The Netherlands
             \and
             INAF - Astronomical Observatory of Padova, Vicolo dell'Osservatorio 5, IT-35122 Padova, Italy
             \and
             Hamburger Sternwarte, Universit\"at Hamburg, Gojenbergsweg 112, D-21029 Hamburg, Germany
             \and
             ASTRON, Netherlands Institute for Radio Astronomy, Oude Hoogeveensedijk 4, 7991 PD, Dwingeloo, The Netherlands
             \\
\email{luca.bruno4@unibo.it}
}


 
  \abstract
   {Radio halos and relics are Mpc-scale diffuse radio sources in galaxy clusters, with a steep spectral index $\alpha>1$ (defined as $S\propto \nu^{-\alpha}$). It has been proposed that halos and relics arise from particle acceleration induced by turbulence and weak shocks, injected in the intracluster  medium (ICM) during mergers.}
   {MACS J1149.5+2223 (MACS J1149) is a high redshift ($z=0.544$) galaxy cluster possibly hosting a radio halo and a relic. We analysed LOw Frequency Array (LOFAR), Giant Metrewave Radio Telescope (GMRT), and  Karl G. Jansky Very Large Array (JVLA) radio data at 144, 323, 1500 MHz, and Chandra X-ray data to characterise the thermal and non-thermal properties of the cluster. }
   {We obtained radio images at different frequencies to investigate the spectral properties of the radio halo. We used Chandra X-ray images to constrain the thermal properties of the cluster and to search for discontinuities (due to cold fronts or shock fronts) in the surface brightness of the ICM. By combining radio and X-ray images, we carried out a point-to-point analysis to study the connection between the thermal and non-thermal emission.}
   {We measured a steep spectrum of the halo, which can be described by a power law with $\alpha=1.49\pm 0.12$ between 144 and 1500 MHz. The radio surface brightness distribution across the halo is found to correlate with the X-ray brightness of the ICM. The derived correlation shows a sub-linear slope in the range 0.4 to 0.6. We also report two possible cold fronts in north-east and north-west, but deeper X-ray observations are required to firmly constrain the properties of the upstream emission.}
   {We show that the combination of high redshift, steep radio spectrum, and sub-linear radio-X scaling of the halo rules out hadronic models. An old ($\sim 1 $ Gyr ago) major merger likely induced the formation of the halo through stochastic re-acceleration of relativistic electrons. We suggest that the two possible X-ray discontinuities may actually be part of the same cold front. In this case, the coolest gas pushed towards the north-west might be associated with the cool core of a sub-cluster involved in the major merger. The peculiar orientation of the south-east relic might indicate a different nature of this source and requires further investigation.}

   
\titlerunning{The steep spectrum radio halo in MACS J1149}
\authorrunning{Bruno et al. 2021}
   \maketitle
%

\section{Introduction}

Galaxy clusters are luminous sources in the X-ray band ($L_{\rm X}\sim 10^{43}-10^{45} \; {\rm erg \; s^{-1}}$) due to the thermal bremsstrahlung emission from the intracluster medium (ICM), i.e. the hot (temperature $T\sim 10^7-10^8$ K) and rarefied (electron density number $n_e\sim10^{-2}-10^{-4}$ cm$^{-3}$) gas that fills the whole cluster \citep{sarazin}. A fraction of dynamically disturbed clusters hosts diffuse synchrotron radio emission, which is primarily classified into radio halos and radio relics, and probes that relativistic particles and magnetic fields co-exist with the thermal gas \citep[][for a recent review]{vanweeren19}. Radio halos extend to Mpc-scales, roughly following the ICM distribution, and they exhibit steep radio spectra\footnote{We define the spectral index $\alpha$ through $S\propto \nu^{-\alpha}$, where $S$ is the radio flux density at the frequency $\nu$.}, with $\alpha \gtrsim 1$. Relics are elongated and arc-shaped structures in the peripheral regions of the cluster, characterised by a spectral index gradient towards the cluster centre, and by a high polarisation fraction. 
   
Observations indicate that complex processes activated by the cluster dynamics operate into the ICM and drain energy into relativistic particles and magnetic fields \citep[see][for a review]{brunettijones14}. Radio halos are thought to originate from re-acceleration mechanisms driven by merger-induced turbulence \citep{brunetti01,petrosian01,brunetti&lazarian07,beresnyak13,miniati15,brunetti&lazarian16}. Models where particles are re-accelerated by turbulence during mergers are supported by the connection found between radio halos and the dynamics of clusters \citep{cassano10A,cassano13,kale13,cuciti15,eckert17,birzan19}. The theoretical challenge in this case is understanding how energy is transported from Mpc scales to smaller scales, where particles interact with turbulence and are accelerated to relativistic velocities \citep{brunettijones14}. Hadronic collisions between thermal and cosmic ray protons (CRp) in the ICM can generate secondary electrons (CRe), which may also produce diffuse synchrotron radiation in the form of radio halos \citep{dennison80,blasi&colafrancesco99,dolag&ensslin00,pfrommer08}. The non-detection of gamma-rays from nearby clusters hosting radio halos \citep{brunetti09,brunetti12,ackermann14,zandanel&ando14,ackermann16} and the very steep spectrum observed in some halos \citep{brentjens08,brunetti08,dallacasa2009,macario10,venturi13,wilber18} rule out a significant contribution from this channel, at least in a number of cases. On the other hand, models where secondary particles are re-accelerated by turbulence are still in agreement with the current data \citep{brunetti&blasi05,brunetti&lazarian11,brunetti17,pinzke17}.

Radio relics are thought to originate at weak (Mach number $\mathcal{M} = 2-4$) cosmic shocks generated in the ICM during mergers and accretion of matter \citep[e.g.][]{ensslin98,roettiger99}. This hypothesis is in line with the observed coincidence of radio relics and merger-shocks, i.e. sharp discontinuities in surface brightness and temperature detected via X-ray observations \citep[e.g.][]{bourdin13,botteon16b,akamatsu17,urdampilleta18}. However, the acceleration efficiency of thermal particles at weak shocks is too small to reproduce the radio spectrum and luminosity of a number of relics \citep[e.g.][]{botteon16a,eckert16,hoang17,botteon20A}. For this reason, models assuming shock re-acceleration of clouds of pre-existing relativistic particles  in the ICM have been proposed \citep[e.g.][]{markevitch05,kang12,pinzke13} and they are supported by the connection between relics and old active galactic nuclei (AGN) plasma observed in some cases \citep[e.g.][]{bonafede14,vanweeren17}.

   In this work, we report on our study of the distant (redshift $z=0.544$) galaxy cluster MACS J1149.5+2223 (hereafter MACS J1149) by means of multifrequency radio and X-ray data. Previous studies of this cluster suggested the presence of two misaligned relics and a candidate radio halo with a very steep spectrum \citep{bonafede12}. Very recently, new results have been published by \cite{giovannini20}, who revised the nature of the candidate west relic and classified it as a radio galaxy. Distant radio halos with steep spectra are particularly important, since turbulent re-acceleration models predict that their number should increase at higher redshift \citep{cassano06,cassano10}, mainly due to stronger inverse Compton radiative losses with the cosmic microwave background (CMB) photons. We present new LOw Frequency ARray (LOFAR) radio data at 144 MHz and Karl G. Jansky Very Large Array (JVLA) data in L-band (1-2 GHz), that we combined with archival Giant Metrewave Radio Telescope (GMRT) data at 323 MHz. In addition to the radio data, we analysed archival Chandra X-ray observations, aiming at searching for surface brightness discontinuities and possible correlations between the thermal and non-thermal emission of the cluster. 
   
   The paper is organised as follows: in Sect. 2 we briefly describe the MACS J1149 galaxy cluster; in Sect. 3 we present the X-ray and radio data and summarise the data reduction and imaging processes; in Sect. 4 we show the results of our analysis; in Sect. 5 we discuss the origin of the halo; in Sect. 6 we summarise our work and give our conclusions.

   Throughout this paper we adopt a $\Lambda$CDM cosmology with $H_0=70\;\mathrm{km\; s^{-1}\; Mpc^{-1}}$, $\Omega_{\rm M}=0.27$ and $\Omega_{\rm \Lambda}=0.73$. At the cluster redshift, the luminosity distance is $D_{\rm L}=3170$ Mpc and $1''=6.4$ kpc.

\section{The galaxy cluster MACS J1149}  
  
  MACS J1149 (RA$_{\rm J2000} \; 11^h49^{m}35^s$, Dec$_{\rm J2000} \;  22^o24'11''$) is a massive\footnote{$M_{500}$ is obtained from $R_{500}$, which is defined as the radius enclosing $500\rho_{\rm c}(z)$, where $\rho_{\rm c}(z)$ is the critical density of the Universe at a given redshift.} \citep[$M_{500}=(10.4\pm0.5)\times 10^{14} \; {\rm M_\odot}$;][]{planckcollaboration16} galaxy cluster, first discovered by the MAssive Cluster Survey \citep[MACS;][]{ebeling01} in the X-ray band. 
  
  Previous lensing and spectroscopy studies revealed the very complex structure of this cluster, with at least three dark matter sub-halos \citep{smith09,golovich16,golovich19}. The two most massive sub-clusters are located north-west and south-east, with respect to the main cluster, at a projected distance of $\sim 1$ Mpc \citep{golovich16}. The third less massive sub-cluster is located $\sim 300$ kpc north from the first one (see Fig. \ref{mappX2}).  
  
  As shown by deep Chandra X-ray observations \citep{ogrean16}, the ICM has a globally high temperature of $kT\sim 10$ keV and is elongated along the NW-SE axis, although the main X-ray concentration is associated with the most massive NW sub-cluster only. The presence of many massive components, the elongated morphology, and the absence of a central cool core concentration are indications of a disturbed state of the cluster. This is further confirmed by the concentration ($c=0.120\pm0.003$) and centroid-shift ($w=0.017 \pm 0.001$) parameters estimated by \cite{yuan&han20}, which locate MACS J1149 among the disturbed clusters in the $c-w$ bimodal distribution reported by \cite{cassano10A}. \cite{ogrean16} also reported the presence of a X-ray surface brightness discontinuity, likely being a cold front, located in the N-NE part of the cluster. 
  
  In the radio band, by using VLA and GMRT data, \cite{bonafede12} found evidence of diffuse emission, with two possible misaligned relics at south-east and west, and a candidate steep spectrum halo, with $\alpha \sim 2$, in line with the complex dynamics of the system. During the preparation of this work, \cite{giovannini20} first presented some of the new JVLA observations of the cluster that we use in our analysis (see Sect. \ref{datiJVLA} for details). They revised the nature of the candidate west relic and classify it as a radio galaxy, and suggested that the south-east relic might be instead a filamentary structure, whose major axis is parallel to the merger one. Throughout the paper, we will refer to the SE source as the `relic', following the original classification proposed by \cite{bonafede12}.

\section{Observations and data reduction}
      \begin{table}
      \centering
   	\caption[]{Details of the Chandra X-ray data analysed in this work (PI: C. Jones).}
   	\label{datiX}
   	\begin{tabular}{cccc}
   	\hline
   	\noalign{\smallskip}
   	ObsID &  CCDs  &  Observation date &    Clean time \\
   	&  &  & (ks) \\
   	\noalign{\smallskip}
   	\hline
   	\noalign{\smallskip}
   	16238 & I0,I1,I2,I3  & 09-Feb.-2015 &   30.4   \\
   	16239 & I0,I1,I2,I3,S2 & 17-Jan.-2015 &   48.8   \\
   	16306 & I0,I1,I2,I3,S2 & 05-Feb.-2014 &   68.7   \\
   	16582 & I0,I1,I2,I3,S2 & 08-Feb.-2014 &   17.6   \\
   	17595 & I0,I1,I2,I3  & 18-Feb.-2015 &   57.1   \\
   	17596 & I0,I1,I2,I3  & 10-Feb.-2015 &   65.2   \\

   	\noalign{\smallskip}
   	\hline
   	\end{tabular}
   \end{table}

   \begin{table*}
   	\caption[]{Details of the LOFAR (LoTSS, pointing name: P177+22), GMRT (PI: M. Br\"uggen, project code: 20\_066), and JVLA (PI: A. Bonafede, project code: 13A-056; PI: R. J. van Weeren, project code: 14B-205) radio data analysed in this work.}
   	\label{datiRADIO}
   	\begin{tabular}{cccccc}
   	\hline
   	\noalign{\smallskip}
   	Instrument &  Configuration  &  Central frequency &  Frequency coverage &  Observation date & On-source time \\
   	&  & (MHz) & (MHz) & & (h) \\
   	\noalign{\smallskip}
  	\hline
   	\noalign{\smallskip}
   	LOFAR & Dual-inner-HBA  & 144 & 120-168 & 04-May-2017 & 8.0     \\
    GMRT & -  & 323 & 305-340 & 27-Jun.-2011, 16-Aug.-2011 & 1.0+4.0     \\
    JVLA & B  & 1500 & 1000-2000 & 17-Nov.-2013, 7-Apr.-2015 & 2.0+3.5     \\
   	JVLA & C  & 1500 & 1000-2000 & 24-Jun.-2013 & 2.5     \\
   	JVLA & D  & 1500 & 1000-2000 & 14-Feb.-2013, 4-Mar.-2013 & 0.75+0.75     \\
   	\noalign{\smallskip}
   	\hline
   	\end{tabular}
   \end{table*}
  
 \begin{table}
      \centering
   	\caption[]{Summary of the parameters adopted to produce the radio maps discussed in the paper. Column 2, 3, and 4 indicate the minimum baseline ($B_{\rm min}$), the robust parameter of the Briggs weighting, and the tapering used, respectively. Column 5 reports the reached resolution ($\theta$).}
   	\label{mapperadioweightscheme}
   	\begin{tabular}{ccccc}
   	\hline
   	\noalign{\smallskip}
   	Dataset & $B_{\rm min}$ & Robust & Taper & $\theta$ \\
   	\noalign{\smallskip}
   	\hline
   	\noalign{\smallskip}
   	LOFAR & $80\lambda$  & $-0.5$ & -  & $10''\times6''$  \\
   	LOFAR & $80\lambda$ & $0$ &  $5''$  & $19''\times15''$  \\
   	${\rm LOFAR^{*}}$ & $170\lambda$ & $0$ & $10''$  & $27''\times21''$  \\
   	${\rm GMRT^{*}}$ & $170\lambda$  & $0$ & $10''$   & $20''\times14''$ \\
   	${\rm JVLA_{\rm [B+C+D]}}$ & - & $0$ & -   & $5''\times4''$ \\
   	${\rm JVLA^{*}_{\rm [D]}}$ & $170\lambda$  & $0$ &  -   & $29''\times28''$  \\

   	\noalign{\smallskip}
   	\hline
   	\end{tabular}
   		\begin{tablenotes}
\item    {\small \textbf{Notes}. $^*$These maps were smoothed with a $30''\times30''$ beam and used to measure the flux density of the halo. } 
 \end{tablenotes}
   \end{table}

In this section we present the X-ray and radio data, summarised in Table \ref{datiX} and Table \ref{datiRADIO}, respectively, and briefly describe the reduction processes. 

\subsection{Chandra X-ray data}
\label{chandrareduction}

We analysed archival Chandra X-ray data \citep[first presented by][]{ogrean16} of MACS J1149, observed in VFAINT mode with ACIS-I. We reprocessed the six considered observations using {\ttfamily CIAO} v. 4.12, with {\ttfamily CALDB} v. 4.9.0. We extracted light curves from the S2 chip (when available) or from one of the I-chips (free of the target emission) in order to filter out soft proton flares with the $lc-clean$ algorithm, which left $287.8$ ks of total clean time. 

We combined each point spread function (PSF) and exposure maps to produce a single exposure-corrected PSF map. This one was used as input in the $wavdetect$ task to find point sources, that we then inspected by eye. Spurious sources were excluded and the confirmed ones were excised.    

The X-ray analysis of galaxy clusters requires a careful treatment of the background, which has various origins: the local hot bubble (LHB), the galactic halo (GH), the unresolved X-ray sources (CXB), and the particle background (NXB). In particular, we modelled the LHB and the GH with a single 
absorbed thermal plasma (APEC model) of $kT=0.20$ keV, and the CXB with an absorbed power law of photon index $\Gamma=1.45$; the NXB was obtained by fitting the single spectra extracted from the reprojected stowed background observations and combining a power law, a decaying exponential and a number of Gaussian profile lines \citep[as described in][]{bartalucci14}. To model the sky components, we extracted the spectra from the observations, in the same regions free of ICM emission and we jointly fitted them with the {\ttfamily XSPEC} v. 12.10.1 package \citep{arnaud96xspec} in the 0.7-11.0 keV band. The NXB emission parameters were kept fixed as obtained by the fit of the single observations. We adopted the Cash statistics \citep[Cstat;][]{cash79} and fixed the (atomic + molecular) hydrogen column density in the direction of the cluster to $n_{\rm H}=2.06\times 10^{20}\; {\rm cm^{-2}}$ \citep{willingale13}. In the following, errors on fitted parameters are quoted at the $68 \%$ confidence level.

\subsection{LOFAR radio data}

Observations of the LOFAR Two-meter Sky Survey \citep[LoTSS;][]{shimwell17,shimwell19LOTSS} are particularly suitable for the detection of extended and low brightness diffuse emission with steep spectrum, such as radio halos and radio relics. MACS J1149 is located at 21 arcmin from the centre of the LoTSS pointing P177+22, that was observed for 8 hours on 4th May 2017 using the High Band Antennas (HBA) Dutch array operating in the 120-168 MHz frequency range. The source 3C 295 was used as flux density scale calibrator.

Data were processed by means of the LOFAR Surveys Key Science Project reduction pipeline v2.2\footnote{\url{https://github.com/mhardcastle/ddf-pipeline}} \citep{shimwell19LOTSS,tasse20}, which performs both direction-independent and direction-dependent calibration using {\ttfamily PREFACTOR} \citep{offringa13,vanweerenDDFACET16,williams16,degasperin19} and {\ttfamily KillMS} \citep{tasse2014b,tasse14a,smirnov15} and delivers images at $6''$ and $20''$ resolution of the full LOFAR field of view using {\ttfamily DDFacet}  \citep{tasse18}. In order to improve the image quality towards MACS J1149, we used the models produced by the pipeline to subtract the sources outside a region of sizes $27'\times27'$ centered on the target from the uv-data. We used this 'extracted' dataset to perform additional cycles of amplitude and phase self-calibration \citep[see details in][]{vanweeren20} using {\ttfamily WSClean} v. 2.7 \citep{offringa14,offringa17}. We then checked for possible systematics in the flux density scale by comparing compact discrete sources in the field with the TIFR GMRT Sky Survey Alternative Data Release \citep[TGSS-ADR;][]{intema17}, finding no significant offset (typical offsets are of the order of $3\%$, with $10\%$ for one source).

\subsection{GMRT radio data}
We retrieved archival GMRT data \citep[first presented by][]{bonafede12} in the 305-340 MHz band, for a total bandwidth of 32 MHz, split into 256 channels. Two observations of 1 and 4 hours on-source time are available, and the sources 3C286 and 3C147 were used as absolute flux density scale calibrators. We reprocessed them independently via the Source Peeling and Atmospheric Modeling ({\ttfamily SPAM}) automated pipeline \citep{intema09}, which corrects for ionospheric effects, and remove direction-dependent gain errors, improving the noise and fidelity of the images. Strong sources in the field of view are used to derive directional-dependent gains and fit a phase-screen over the entire field of view. Finally, images are corrected for the system temperature variations between the calibrator and the target.

During the imaging steps, we noticed that the flux density of the point sources in the field were different for the two observations, independently of the adopted primary calibrator. In order to check for the spectra of the sources, we measured their flux density at 144 (LOFAR), 610 (GMRT), 1500 (JVLA, L band), and 2800 (JVLA, S band) MHz\footnote{Details of the 610 and 2800 MHz data are not reported, as we used them only to check the flux density scale of the point sources in the field. The halo is too faint and is undetected at 2800 MHz, and it is not fully recovered at 610 MHz due to the insufficient depth and missing short baselines of these data.}. The expected power laws are well fitted in the case of the shallower GMRT data, whereas the deeper ones appear systematically biased low. The logs of the 4 hours observation report of high wind speeds, which could have affected some scans and system gains. To account for this problem, we corrected the deeper images by multiplying them for a factor of 1.23, i.e. the mean ratio of the source flux densities in the shallow and deep maps.  

\subsection{JVLA radio data}
\label{datiJVLA}
MACS J1149 was observed with the JVLA at 1-2 GHz (L-band) in B, C, and D configurations, for 5.5, 2.5 , and 1.5 hours on-source time, respectively (see Table \ref{datiRADIO} for details), with 3C286 as flux density calibrator and J1150+2417 as phase calibrator. The data were recorded with 16 spectral windows, each divided into 64 channels. It is worth to mention that, during the preparation of this work, data collected in 2013 (except for the observation in D array of the 4th March) were first published by \cite{giovannini20}.
Data reduction was carried out with the National Radio Astronomy Observatory (NRAO) Common Astronomy Software Applications \citep[{\ttfamily CASA};][]{mcmullincasapaper07} v. 4.7. We removed Radio Frequency Interference (RFI) by means of the \verb|AOFlagger| software \citep{offringa10,offringa12}. The data from B, C, and D configurations were calibrated independently, by performing the standard calibration procedure, which includes the initial phase, bandpass, and gain calibrations. In addition, we performed two rounds of phase-only self-calibration and two rounds of amplitude and phase self-calibration on the individual datasets. Finally, we combined the B, C, and D configuration data to create deep images.

\subsection{Radio imaging and source subtraction}
\label{wscleansub}
The imaging process was carried out for all the datasets with {\ttfamily WSClean} v. 2.7, which is designed for wide-field, multifrequency, and multiscale synthesis. All images are corrected for the primary beam attenuation. For each dataset, we produced maps with different baseline weightings to study the radio emission of the target on various spatial scales. Short baselines are sensitive to the extended emission, whereas long baselines provide a higher resolution. A compromise between sensitivity and resolution can be reached with the robust parameter intermediate weighting \citep{briggs95} and a tapering of the baselines. 

The weighting schemes adopted for the maps discussed in the next sections are summarised in Table \ref{mapperadioweightscheme}. In order to properly image the diffuse emission and consistently compare the LOFAR, GMRT, and JVLA (D array) maps, we adopted a common minimum baseline of $170\lambda$ (which gives a maximum recoverable scale of $\sim 20'$, corresponding to $\sim 8$ Mpc), a {\ttfamily briggs} weighting with robust 0, and a tapering of the long baselines ($10''$), then we smoothed these maps with a $30''\times30''$ beam. In addition, we produced an intermediate resolution ($19''\times15''$) LOFAR map, that we used for subsequent analysis (see Sect. \ref{sectionTnT} and \ref{sectionadronici}). 

In order to measure the flux density of the halo, we have to remove the contribution of the sources embedded in it. For LOFAR and GMRT datasets, we selected the sources by imaging the data at high resolution, excluding the shortest baselines ($<4{\rm k}\lambda$, corresponding to maximum angular scales of $\sim 50''$), then we subtracted the clean components in the model image from the uv-data. We tested the consistency of this process by comparing the flux density of the halo with that obtained by algebraically subtracting the flux density of the embedded sources from the total (halo $+$ sources) one, finding a good agreement (within $3\%$ and $5\%$ of the flux density for LOFAR and GMRT, respectively). Subtracting the clean components of the discrete sources is not sufficiently accurate with the JVLA datasets, given the very low flux density of the halo. The low sensitivity to extended sources of the B array and the insufficient depth of the C array data, do not allow us to accurately recover the halo emission. On the other hand, the low resolution of the D array makes it difficult to obtain a good model of both point and extended sources. However, we can use the sensitive high resolution images to identify faint compact sources in the halo region, which contribute to the total flux density when measured from lower resolution images. Therefore, we accurately measured the flux density of point sources in B and C configurations at high resolution, and the extended sources in D configuration at low resolution, then we subtracted their contribution out of the total (halo + sources) flux density obtained in D array. The model subtraction from the uv-data bias the net flux density of the halo at 1.5 GHz high, but the algebraic subtraction might bias it low; to take into account this effect, we safely consider the $20\%$ of the flux density as the uncertainty on the subtraction.

In the following, the position angle ($P.A.$) of the beam is measured north-eastwards. The uncertainties $\Delta S$ on the reported flux densities are given by:
\begin{equation}
\Delta S= \sqrt{ \left( \sigma \cdot \sqrt{N_{\rm beam}} \right)^2 + \left(  \xi_{\rm cal} \cdot S \right) ^2+\left(  \xi_{\rm sub} \cdot S \right) ^2}
\label{erroronflux}
\end{equation}
where $\sigma$ is the RMS noise of the map, $N_{\rm beam}$ is the number of independent beams within the considered region, $\xi_{\rm cal}$ is the calibration error, and $\xi_{\rm sub}$ is the uncertainty of the source subtraction. As mentioned, we assumed $\xi_{\rm sub}=3\%, \; 5\%, \; {\rm and} \; 20\%$ for LOFAR, GMRT, and JVLA, respectively. We adopted $\xi_{\rm cal}=10\%$ for LOFAR \citep[lower than the $20\%$ reported in][given the low-significance offset found with respect to the TGSS image]{shimwell19LOTSS}, $\xi_{\rm cal}=10\%$ for GMRT \citep{chandra04}, and $\xi_{\rm cal}=5\%$ for JVLA \citep{perley&butler13}.

\section{Results}

In this section we first show the results of the X-ray and radio data analysis, then we investigate the connection between the thermal and non-thermal emission.  

\subsection{Global X-ray properties}
\begin{figure*}
	\centering
	\includegraphics[height=7cm,width=7.7cm]{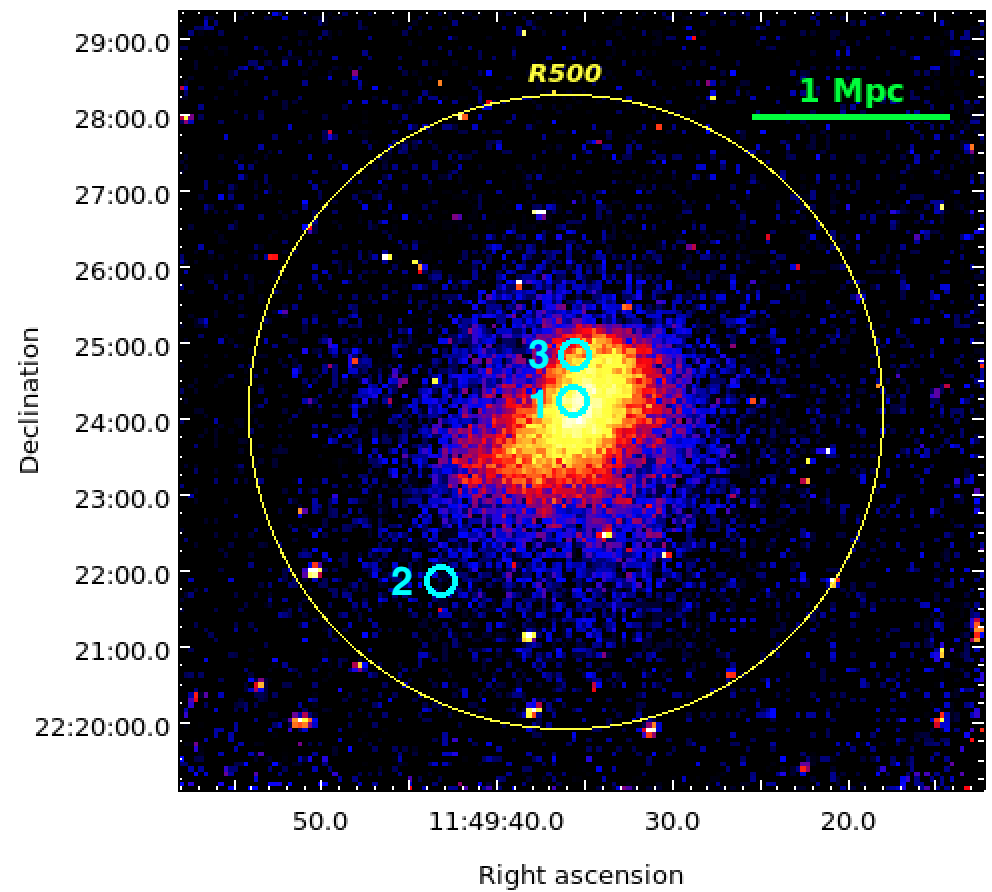}
	\includegraphics[height=6.8cm,width=8.1cm]{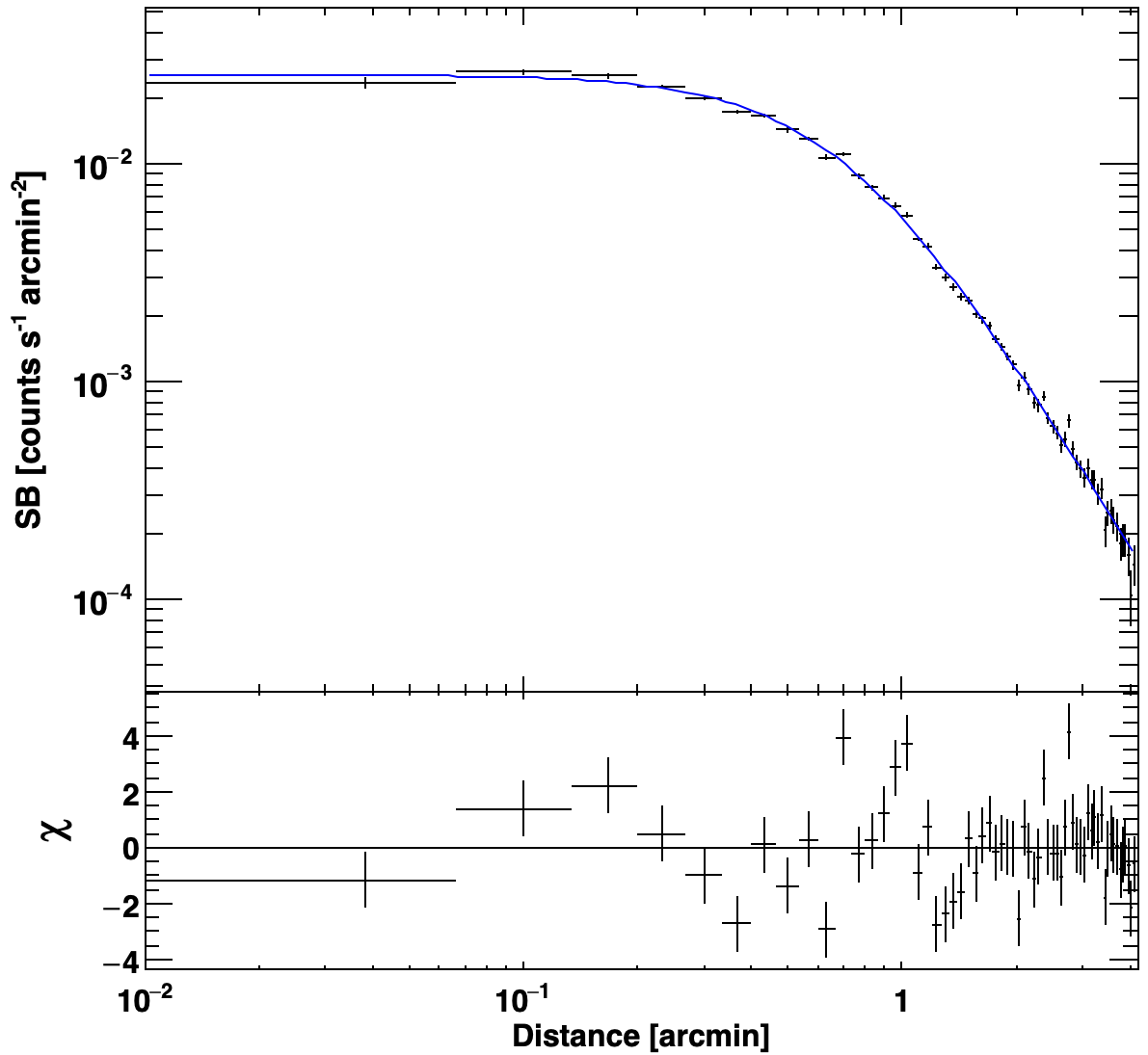}
	\smallskip
	
	\caption{{\it Left:} MACS J1149 X-ray map obtained with Chandra in the 0.5-3 keV band ($1 \; {\rm pix}=4''$). The region within $R_{500}$ is indicated in yellow and cyan circles label the location of the three sub-clusters, as reported by \cite{golovich16}. The cluster is elongated along NW-SE. Sub-cluster 1 is associated with the main X-ray concentration, whereas no clumps associated with sub-cluster 2 are clearly visible. {\it Right:} X-ray surface brightness profile. Data are fitted by a $\beta$-model (blue line) with $\beta=0.66\pm0.01$ and $r_c=288\pm8 \; {\rm kpc} $.}
	\label{mappX2}%
\end{figure*}

\begin{figure}
	\centering
	\includegraphics[height=6cm,width=7.5cm]{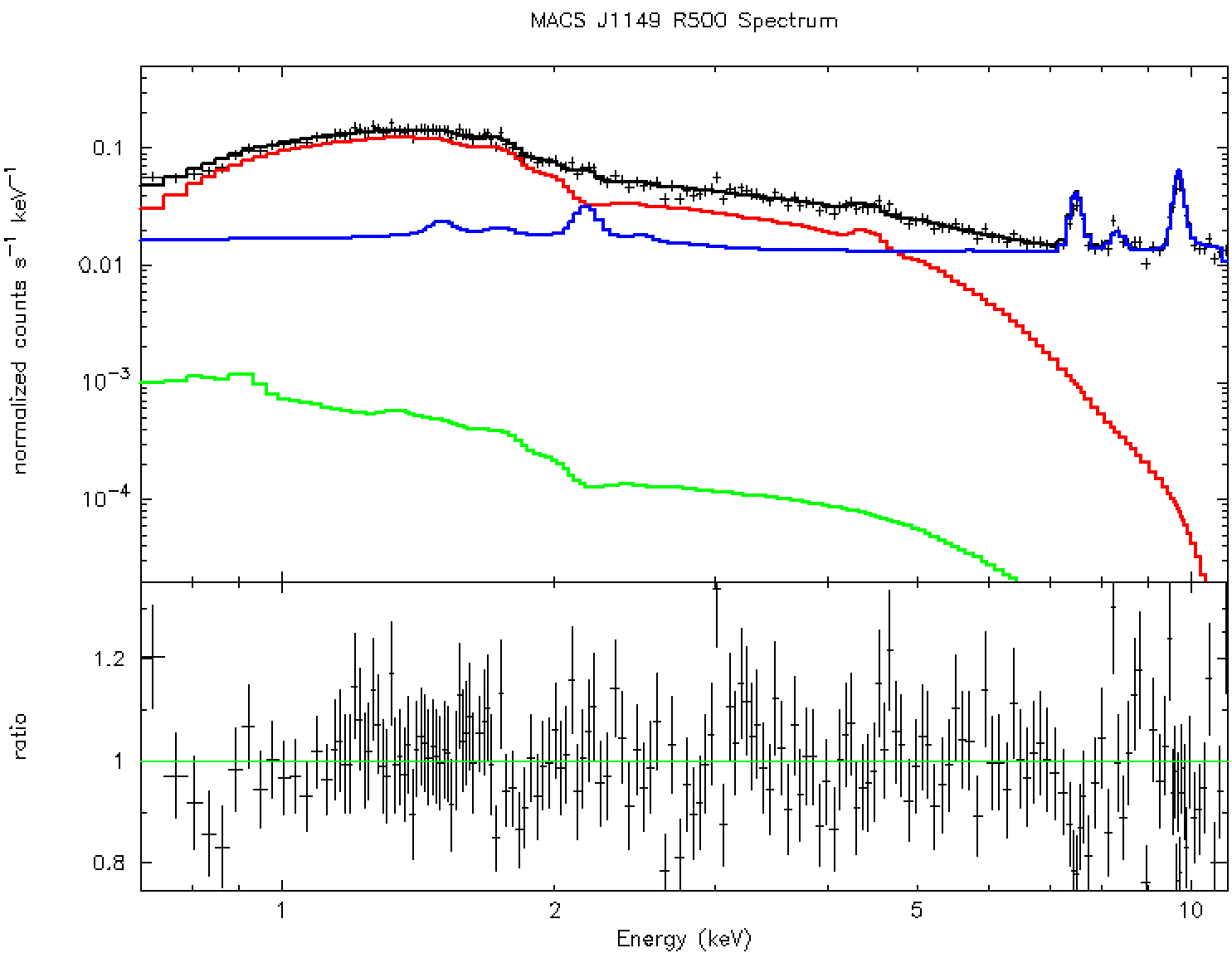} 
	\caption{MACS J1149 X-ray spectrum within $R_{500}$. Data and the best fit model are shown in black. The red, green, and blue curves indicate the cluster emission, the sky background, and the particle background, respectively. The bottom panel shows the ratio of the data to the model. Fits were performed simultaneously for all the observations, but we report only one spectrum for a clearer inspection of the plot. We obtained $kT=10.5\pm0.3$ keV and $Z=0.23\pm0.05\; \mathrm{Z_{\odot}}$. }
	\label{spettroR500}%
\end{figure}

In Fig. \ref{mappX2} (left) we present the exposure-corrected map of MACS J1149 in the 0.5-3 keV band, obtained by merging the six Chandra observations listed in Table \ref{datiX}. The cyan circles indicate the locations of the three sub-clusters, as reported by \cite{golovich16}; they have virial masses of $M_{\rm 1} \sim M_{\rm 2} \sim 10^{15} \; {\rm M_{\odot}}$, $M_{\rm 3} \sim 10^{14} \; {\rm M_{\odot}}$, respectively. The ICM appears elongated along the NW-SE axis, where sub-clusters 1 and 2 lie. Even though sub-clusters 1 and 2 have similar masses, only the most massive sub-cluster 1 is associated with the main X-ray concentration (the depth of the data would be sufficient to detect X-ray clumps associated with sub-cluster 2, if any).
 
In the right panel of Fig. \ref{mappX2} we show the surface brightness profile within a circle (centered on the pixel coincident with the X-ray peak) of radius $R_{500}=1.6$ Mpc, extracted and fitted by means of the {\ttfamily PROFFIT} v. 1.5 code \citep{eckert11}. The background was assumed to be constant. It was estimated in a source-free region and subtracted, whereas the data were fitted with a simple beta-profile under the assumptions of a isothermal gas and a spherical geometry \citep{fusco-femiano&cavaliere76}:
\begin{equation}
   n_{\rm th}(r)=n_{\rm th,0}\left[1+\left(\frac{r}{r_{\rm c}}\right)^2 \right]^{-\frac{3}{2}\beta}
\label{betamodeleq}
\end{equation}
where $n_{\rm th}$ is the thermal particle density number, $r_{\rm c}$ is the core radius, and the index $\beta$ represents the ratio of the kinetic energy of the galaxies to the thermal energy of the gas. The thermal particle distribution is related to the X-ray surface brightness profile $I_{\rm X}(r)$ as \citep{sarazin}:
\begin{equation}
   I_{\rm X}(r)=\sqrt{\pi}n_{\rm th,0}^2r_{\rm c}\Lambda(T)\frac{\Gamma(3\beta-0.5)}{\Gamma(3\beta)}\left[1+\left(\frac{r}{r_{\rm c}}\right)^2 \right]^{\frac{1}{2}-3\beta}
\label{betamodelSB}
\end{equation}
where $\Lambda(T)\sim2.5\times10^{-27}\sqrt{T} \; {\rm erg \; cm^{3} \; s^{-1}}$ is the cooling function in the case of a plasma at a temperature $T$ emitting through thermal bremsstrahlung  \citep[e.g.][]{gitti&brighenti&mcnamara12}. 
The beta-model has commonly been used to describe the radial profile of a number of galaxy clusters, providing reasonable representations of the data \citep{arnaud09} even with $\chi^2/{\rm dof}\sim 2$ \citep[see e.g. Table 5 reported in][]{eckert11}. We obtained $\beta=0.66\pm0.01$ and $r_{\rm c}=288\pm8 \; {\rm kpc} $, with a $\chi^2/{\rm dof}=2.3$, consistent with the typical values found for merging clusters \citep{jones&forman84}.

Finally, the X-ray spectrum extracted within $R_{500}$ is shown in Fig. \ref{spettroR500} (we jointly fitted all the observations, but we report only one spectrum to avoid confusion in the plot). The particle and sky background components were modelled as described in Section \ref{chandrareduction}, whereas we adopted an absorbed APEC for the ICM emission. We obtained a global temperature of $kT=10.5\pm0.3$ keV and a metallicity of $Z=0.23\pm0.05\; \mathrm{Z_{\odot}}$ ($\mathrm{Cstat/dof}=3530/3076$). The computed flux within $R_{500}$ is $F_\mathrm{ [0.1-2.4 \; keV]}^{500}=(1.67\pm0.01)\times10^{-12} \; \mathrm{erg \; s^{-1} \; cm^{-2}}$, which gives a luminosity of $L_\mathrm{
[0.1-2.4 \; keV]}^{500}=(1.35\pm0.01)\times10^{45} \; \mathrm{erg \; s^{-1}}$, in agreement with the values reported by \cite{ogrean16}.

\subsection{Surface brightness edges}

\begin{figure*}
	\centering
	
	\includegraphics[height=6.5cm,width=8cm]{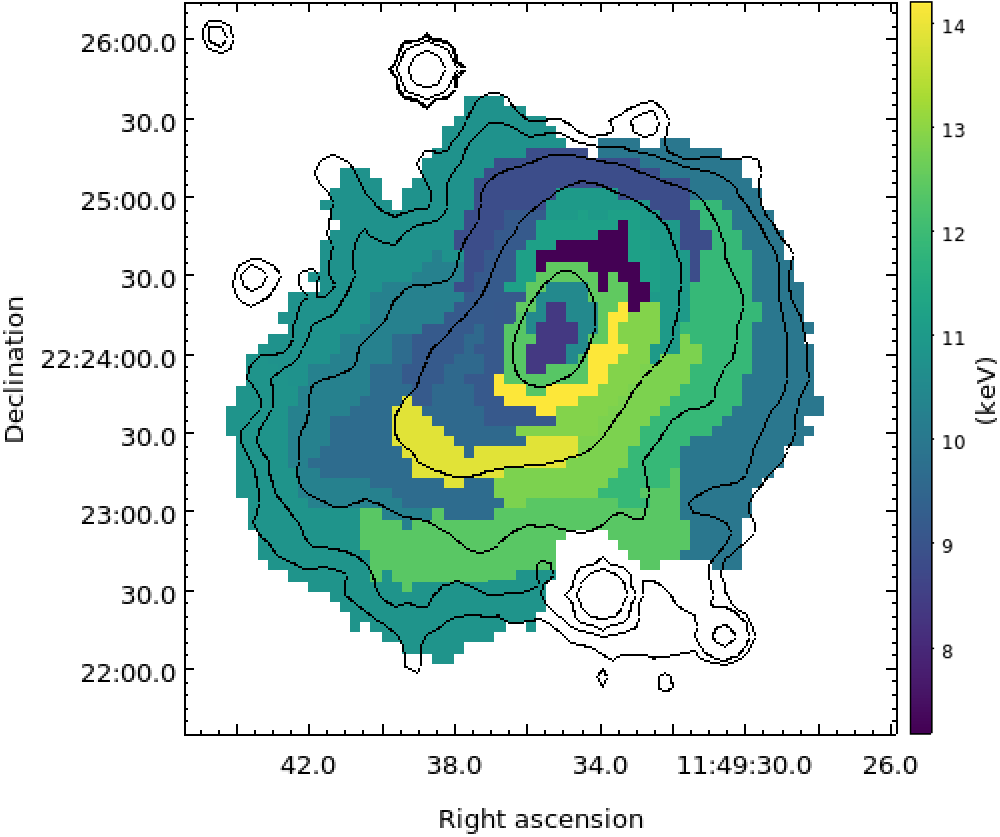} 
	\includegraphics[height=6.5cm,width=8cm]{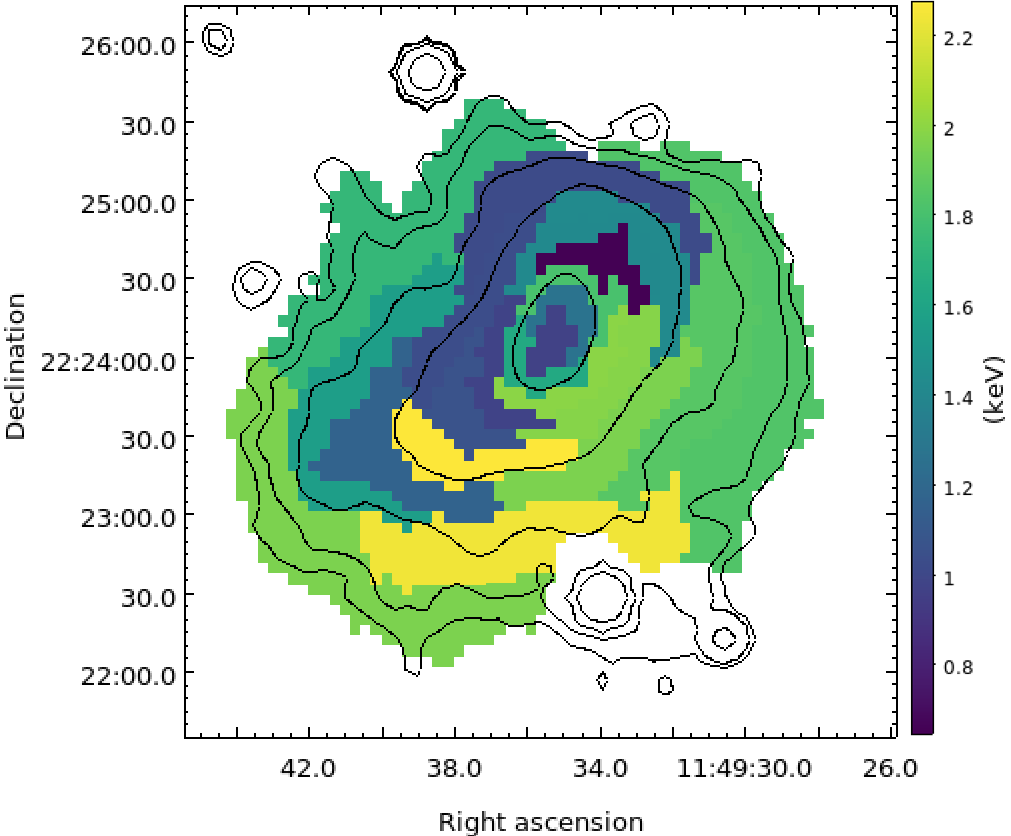}
	\caption{Temperature (left) and associated error (right) maps, reported in keV units. The X-ray surface brightness contours of Fig. \ref{mappX2} are reported in black. The temperature map suggests the presence of a significant lower kT region in NW. }
	\label{mappaT}%
\end{figure*}   

\begin{figure*}
	\centering
	
	\includegraphics[height=6.5cm,width=7.5cm]{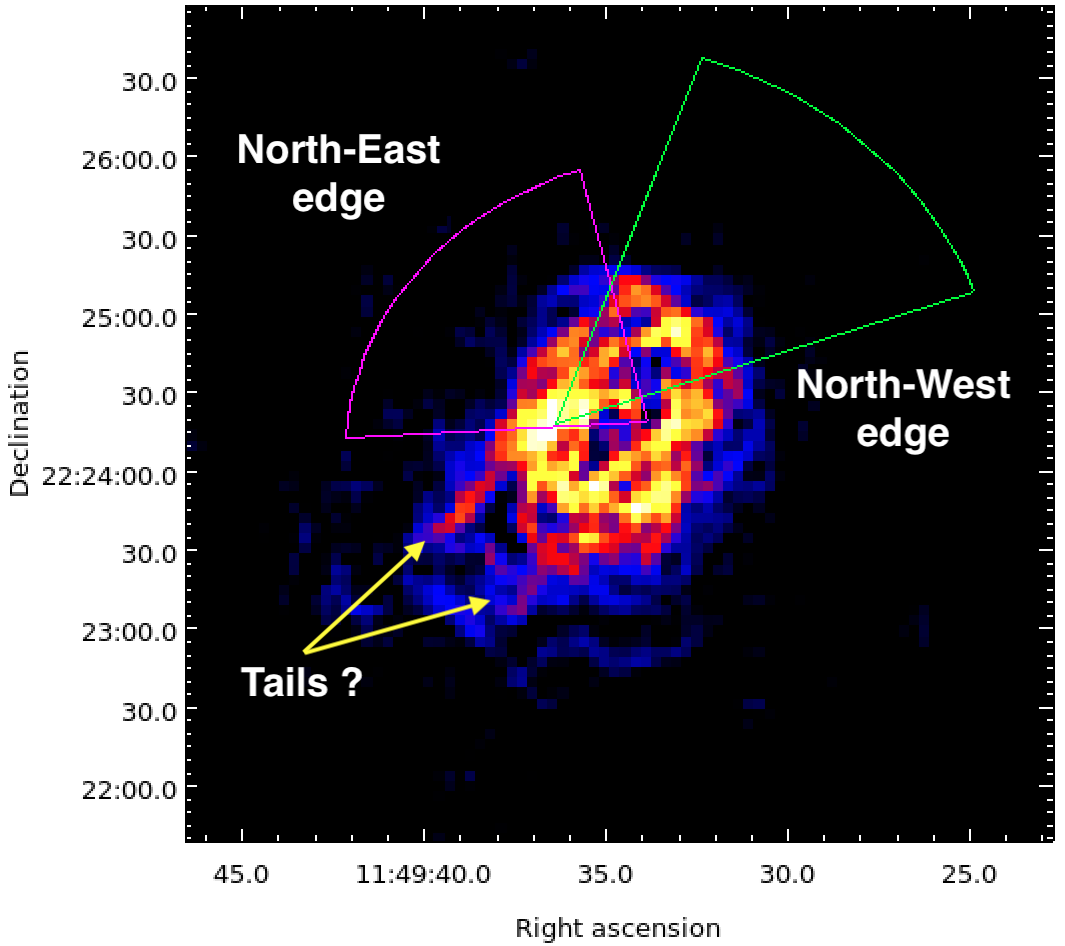} \\
	\includegraphics[height=6cm,width=7.5cm]{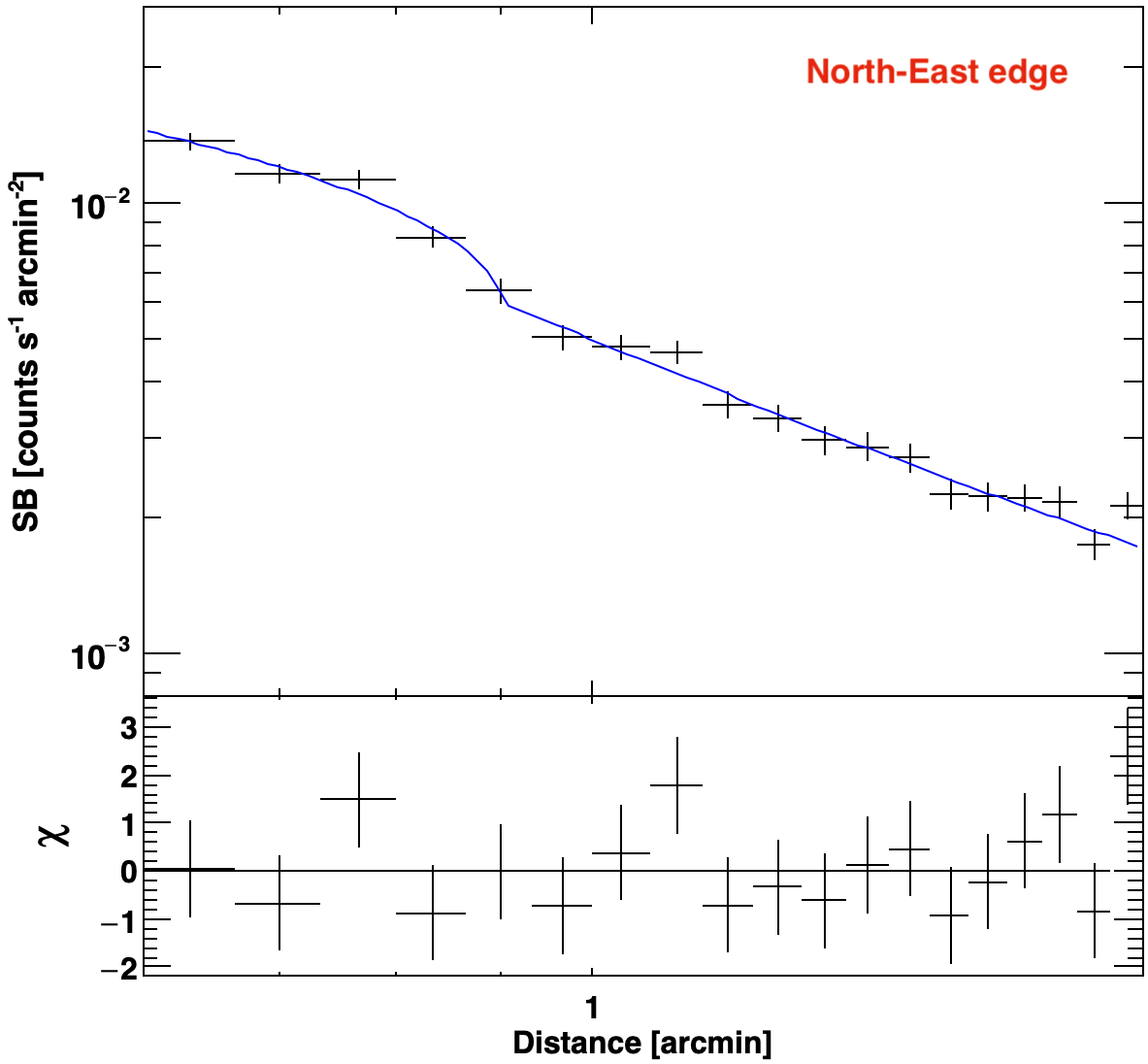}
	\includegraphics[height=6cm,width=7.5cm]{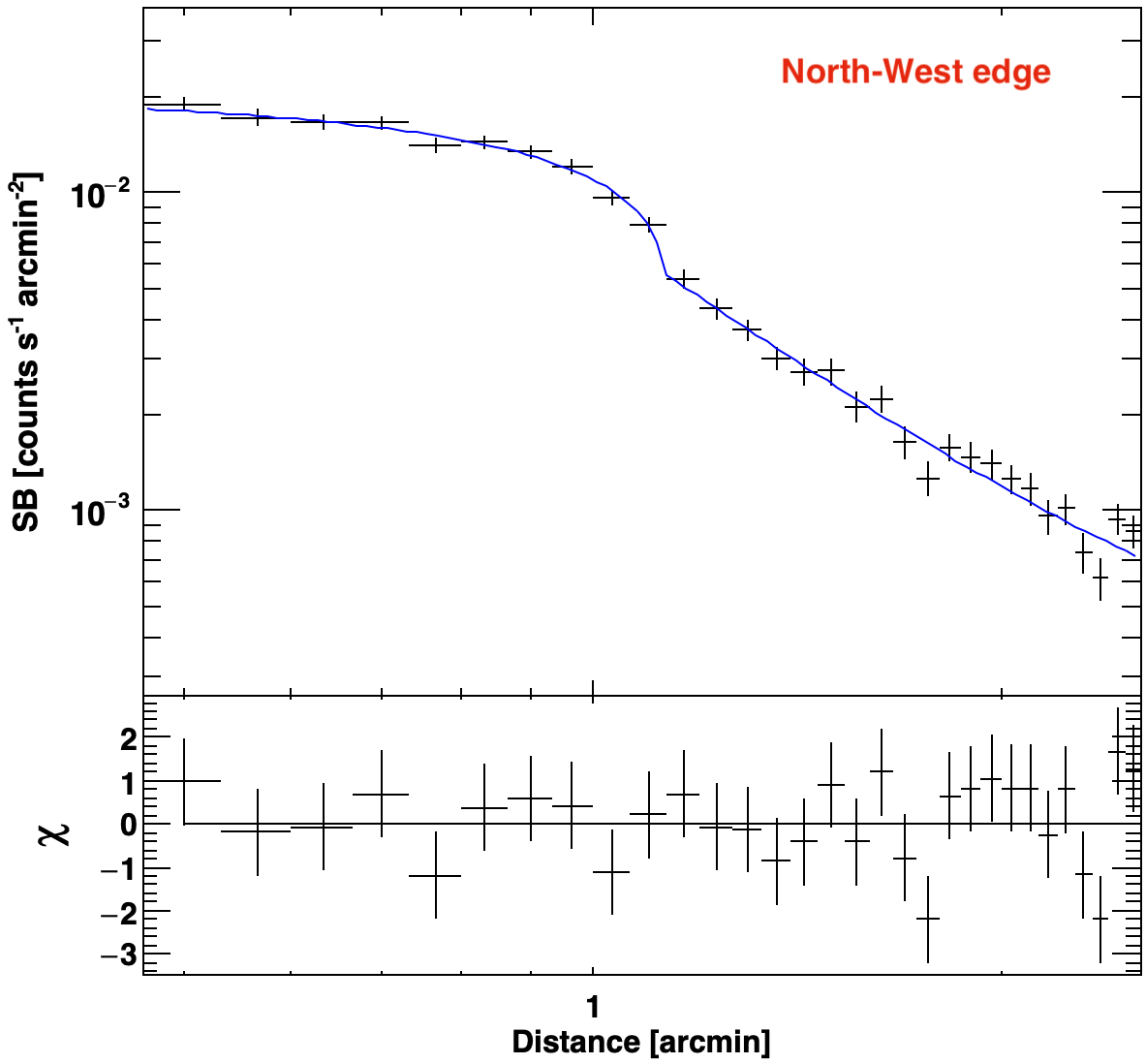} \\
	\includegraphics[height=6.5cm,width=7.5cm]{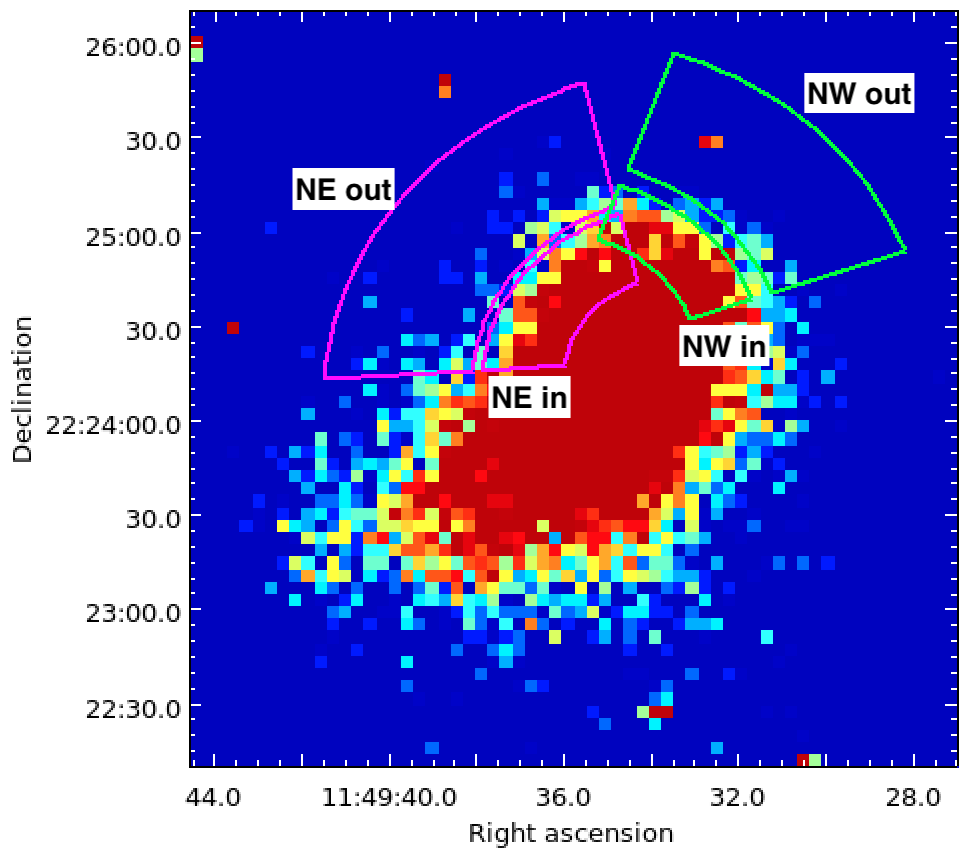}
	\caption{{\it Top:} GGM-filtered X-ray image with $\sigma_{\rm GGM}=1$ pix ($1 \; {\rm pix}=4''$). Two possible surface brightness edges are revealed in the NE and NW. Two filamentary structures (the 'tails') are present at south. {\it Middle left:} Best fit ($\chi^2/{\rm dof}=0.99$) of the NE edge with a broken power law. The density compression factor is $C=1.38\pm 0.11$. {\it Middle right:} Best fit ($\chi^2/{\rm dof}=1.12$) of the NW edge with a broken power law. The density compression factor is $C=1.52\pm 0.11$. A fit with a single power law is rejected ($\chi^2/{\rm dof}=1.80$) and thus not shown. {\it Bottom:} X-ray count image showing the sectors used to extract the spectra upstream and downstream the surface brightness discontinuities. The total source and background counts are $N_{\rm in, NE}\sim2600$, $N_{\rm out, NE}\sim 2300$ in the NE sectors, and $N_{\rm in, NW}\sim1900$, $N_{\rm out, NW}\sim 1300$ in the NW sectors. We do not find significant temperature jumps  ($kT_{\rm in, NE}=9.8^{+0.9}_{-0.9}$ keV and $kT_{\rm out, NE}=10.7^{+1.5}_{-1.1}$ keV; $kT_{\rm in, NW}=9.0^{+1.0}_{-0.8}$ keV and $kT_{\rm out, NW}=11.8^{+2.3}_{-3.1}$ keV).  }
	\label{ggm}%
\end{figure*}

\begin{table}
\centering
	\caption[]{Properties of the surface brightness discontinuities in Fig. \ref{ggm}. Column 3 indicates the distance ($R_{\rm D}$) of the edge from the sector centre. Column 4 reports the density compression factor ($C$). Column 6 lists the temperature in the inner and outer sub-sectors.}
	\label{edgevalues}   
	\begin{tabular}{cccccc}
	\hline
	\noalign{\smallskip}
	Edge & Centre & $R_{\rm D}$ & $C$ & Region  & $kT$ 
	\\  
	 &  & (arcmin) & & & (keV)   \\  
    \hline
	\noalign{\smallskip}
	 &  & &  & inner & $9.8^{+0.9}_{-0.9}$   \\
	NE & $^{11:49:33.8} _{+22:24:18.9}$ & $0.90^{+0.01}_{-0.01}$ & $1.38^{+0.11}_{-0.11}$ &  &  \\
   &  & & & outer & $10.7^{+1.5}_{-1.1}$    \\
   &  & & & & \\
        \hline
	\noalign{\smallskip}
	 &  & &  & inner & $9.0^{+1.0}_{-0.8}$   \\
	NW & $^{11:49:36.3} _{+22:24:17.9}$ & $1.13^{+0.02}_{-0.01}$ & $1.52^{+0.11}_{-0.11}$ &  &  \\
   &  & & & outer & $11.8^{+2.3}_{-3.1}$    \\
   &  & & & & \\
        \hline
	\noalign{\smallskip}
	\end{tabular}
\end{table}

We searched for possible sub-structures in the X-ray surface brightness distribution. As a first step, we used the {\ttfamily CONTBIN} v. 1.6 code \citep{sanders06} to adaptively bin the surface brightness map into regions with a signal-to-noise ratio of ${\rm SNR}=30$, thus ensuring enough source and background counts. Then, we extracted the spectra within those regions and jointly fitted them (the background and metallicity were kept fixed to the previously obtained values) in order to produce the temperature map shown in Fig. \ref{mappaT} (the error map on the right was obtained as the mean of the lower and upper errors). As a general trend, we notice that the eastern regions of the cluster are colder than the western ones \citep[this is also visible in the temperature map shown in][and obtained with a lower ${\rm SNR}=10$]{ogrean16}. Moreover, we observe northwards and southwards a difference in temperature, with respect to the global value, but the error map indicates that only the northern region has a temperature significantly different ($kT=7.2 \pm 0.6$ keV) from that of the surrounding bins. 

 Discontinuities and sub-structures in the surface brightness profile can be enhanced by means of the Gaussian gradient magnitude \citep[GGM;][]{sanders16b,sanders16a} filtering. We tested different $\sigma_{\rm GGM}$ scales (i.e. the width of the derivative of the Gaussian curve): too narrow scales ($<4''$) are usually not efficient in detecting edges in the outskirts of the cluster, due to the low counts, whereas too large scales ($>8''$) could give excessively smoothed images \citep[e.g.][]{botteon18b}. We found that possible discontinuities are better highlighted with $\sigma_{\rm GGM}=1,2 \; {\rm pixels}$ (where 1 pixel correspond to $4''$ in our maps). 

In Fig. \ref{ggm} (top panel) we show the GGM-filtered image with $\sigma_{\rm GGM}=1$ pix. This map reveals two distinct possible edges in NE and NW, instead of the single one reported by \cite{ogrean16}, that partially covers both of them. In particular, the candidate NW edge is coincident with the cold region highlighted by the temperature map. We do not find discontinuities in the south, however we confirm the presence of two filamentary structures, as found with the unsharp mask analysis by \cite{ogrean16}, who referred to them as the X-ray tail (SW) and peak (SE). The GGM-filtering suggests that both of them could be tails, likely associated with the cluster dynamics.

In order to study the two candidate discontinuities, we extracted the surface brightness profiles in the purple and green sectors, and we fitted them with a broken power law model \citep[e.g.][]{owers09}:
\begin{equation}
   n(r)=\begin{cases}
   Cn_0\left( \frac{r}{R_{\rm D}} \right)^{-a} \; {\rm if \; r \le R_{\rm D} }
   \\ \\
   n_0\left( \frac{r}{R_{\rm D}} \right)^{-b} \; {\rm if \; r > R_{\rm D} }
   \end{cases}
\label{brokenpow}
\end{equation}
where $C$ is the density compression factor, $a$ and $b$ are the power law slopes, and $R_{\rm D}$ is the radius of the edge. The fitted profiles are shown in Fig. \ref{ggm} (middle panels) and the obtained values are summarised in Table \ref{edgevalues}. The two discontinuities give similar compression factors, with $C=1.38 \pm 0.11$ for NE, and $C=1.52 \pm 0.11$ for NW. The NE edge is less prominent than the NW one and we also tested a single power law model. The resulting $\chi^2/{\rm dof}$ of the fits are 0.99 and 1.80 with the broken power law and the single power law, respectively, thus we rejected the latter case.

The nature of the edges can be investigated by the sign of the jump in temperature. Therefore, we extracted the spectra in inner (downstream) and outer (upstream) sectors, with respect to the radii of the jumps (see Fig. \ref{ggm}, bottom). We did not find clear indications of temperature discontinuities: in fact we obtained $kT_{\rm in, NE}=9.8^{+0.9}_{-0.9}$ keV and $kT_{\rm out, NE}=10.7^{+1.5}_{-1.1}$ keV for the NE edge, and $kT_{\rm in, NW}=9.0^{+1.0}_{-0.8}$ keV and $kT_{\rm out, NW}=11.8^{+2.3}_{-3.1}$ keV for the NW edge (see Table \ref{edgevalues}). 

Under the assumption of a shock moving outwards from the cluster centre, the Hugoniot-Rankine density jump condition for a mono-atomic thermal gas can be used to compute the corresponding Mach-number:
\begin{equation}
    {\mathcal{M}}=\sqrt{\frac{3C}{4-C}}
\label{hugoniot-rankine1}
\end{equation}
The associated temperature jump is given by:
\begin{equation}
\frac{T_{\rm in}}{T_{\rm out}}=\frac{5\mathcal{M}^4+14\mathcal{M}^2-3}{16\mathcal{M}^2}
\label{hugoniot-rankine2}
\end{equation}
Eq. \ref{hugoniot-rankine1} provides Mach-numbers of ${\mathcal{M}_{\rm NE}}=1.26\pm 0.08$ and ${\mathcal{M}_{\rm NW}}=1.36\pm 0.08$, implying temperature jumps of $\left(T_{\rm in}/T_{\rm out}\right)_{\rm NE}=1.25\pm 0.07$ and $\left(T_{\rm in}/T_{\rm out}\right)_{\rm NW}=1.35\pm 0.08$. In the most favourable scenario, by assuming the upper limits for the downstream temperatures, i.e. $kT_{\rm in, NE}=10.7$ keV and $kT_{\rm in, NW}=10.0$ keV, the corresponding upstream temperatures from Eq. \ref{hugoniot-rankine2} would result $kT_{\rm out, NE}=8.6\pm 0.5$ keV and $kT_{\rm out, NW}=7.4\pm 0.4$ keV. We fixed these temperatures and re-fitted the spectra, but the Cstat increases ($\Delta{\rm Cstat_{\rm NE}=1.9}$, $\Delta{\rm Cstat_{\rm NW}=4.4}$). Lower downstream temperatures would result in even worse fit, thus we conclude that the edges are unlikely to be shock fronts.  

The two discontinuities may be associated with cold fronts. Indeed, the N-NE edge revealed by \cite{ogrean16} was classified as a candidate cold front, which would be one of the most distant cold fronts to date, if confirmed. This N-NE edge is roughly located in the middle between the NE and NW edges we found and has a compression factor consistent with ours. Even though the GGM-filtering is efficient in separating the discontinuities, the low source counts in the upstream sectors do not allow us to confirm whether the NE and NW can be considered as a single or double edge (in this regard, projection effects may play an important role). That said, we note that the temperature map in Fig. \ref{mappaT} suggests similarities with the prototypical case of Abell 3667 \citep{owers09,ichinohe17}, where the coolest gas of the system is pushed at the tip of the cold front, surrounded by the hotter gas. This interpretation would be in line with the merger scenario described by \cite{golovich16} between sub-cluster 1 and sub-cluster 2.

\subsection{Radio maps and properties}
\label{sectionmappe}

\begin{figure*}
	\centering
	\includegraphics[height=8.2cm,width=8.5cm]{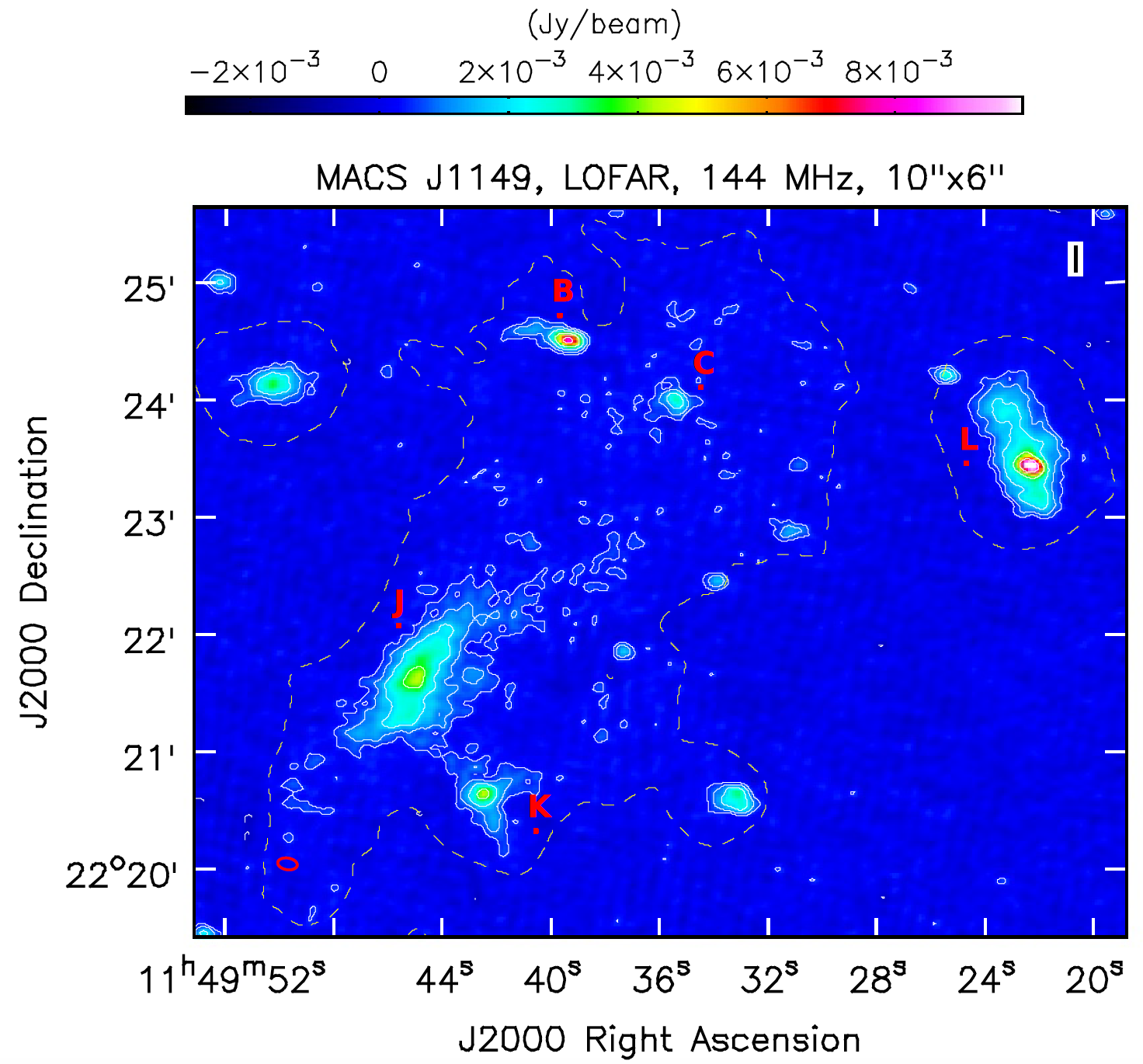}
	\includegraphics[height=8.2cm,width=8.5cm]{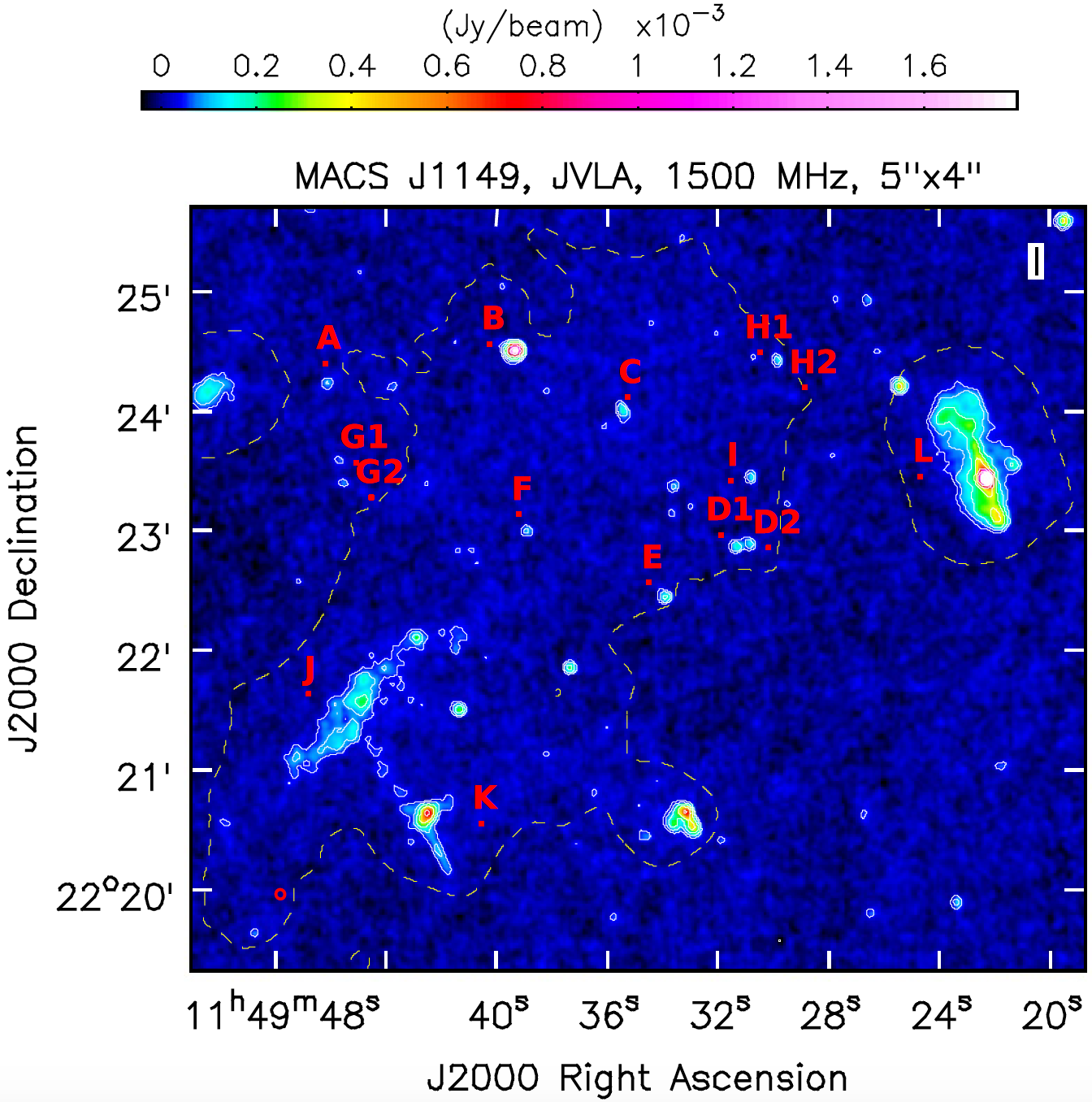}
	\includegraphics[height=6cm,width=8cm]{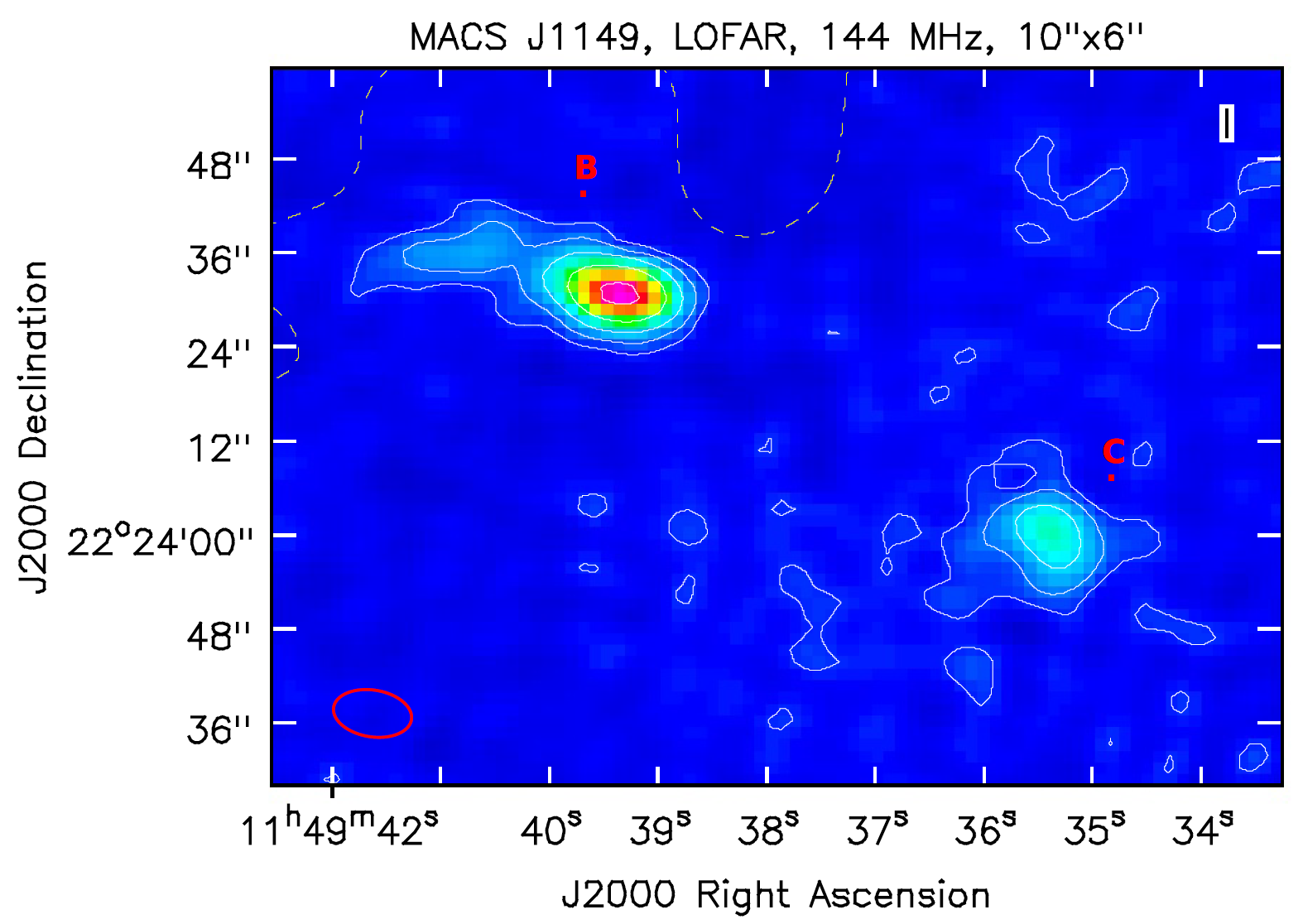}
    \includegraphics[height=5.8cm,width=5.4cm]{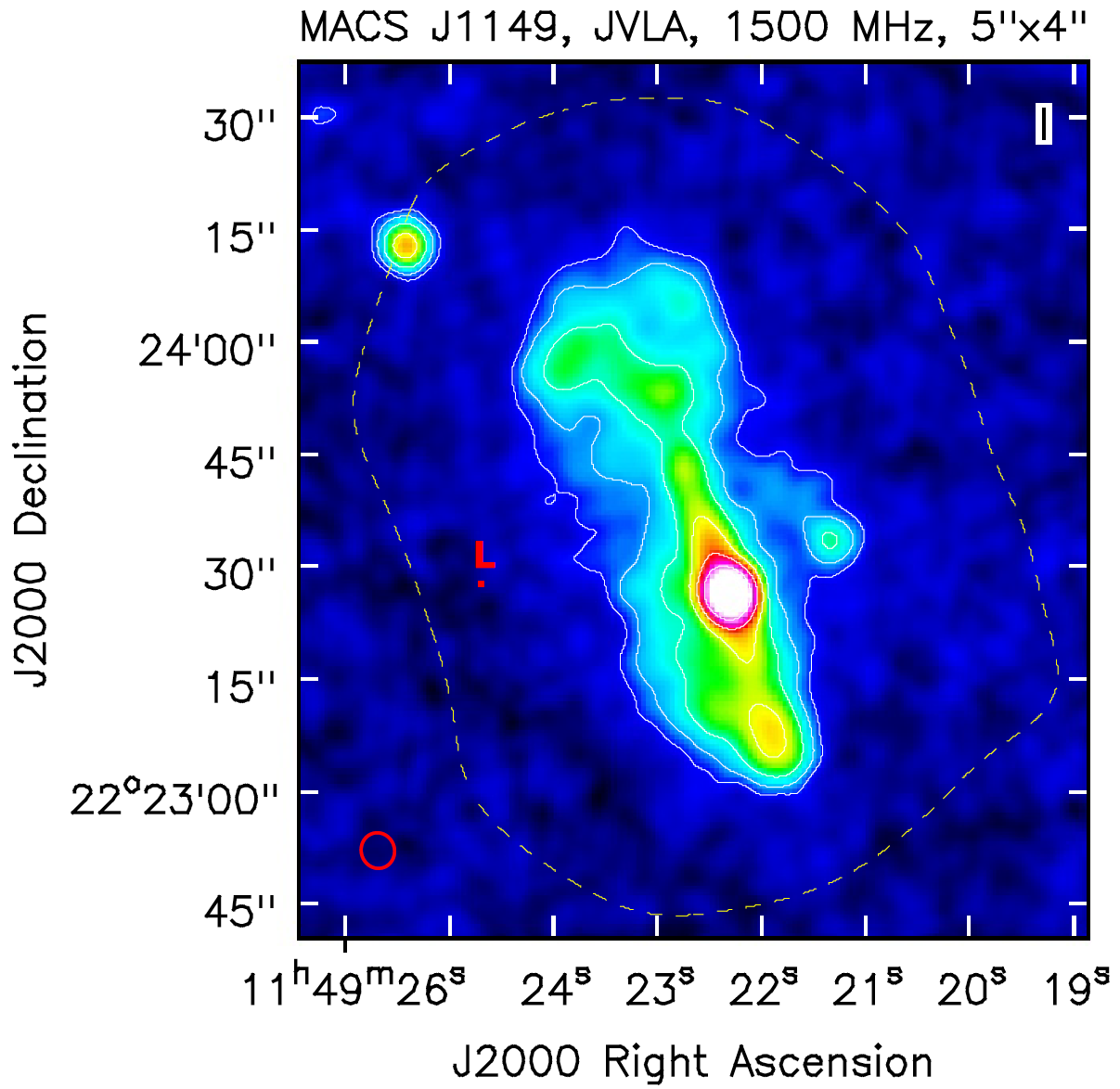}
	\caption{MACS J1149 radio maps at high resolution. In all the panels, the white contour levels are $[\pm3, \;6, \;12, ...]\times \sigma$. The dashed yellow contour indicates the $3\sigma$ (with $\sigma \sim 0.30 \; {\rm mJy \; beam^{-1}}$) level of the $30''\times30''$ resolution LOFAR map after the subtraction of the embedded sources (see also Fig. \ref{mappe}). {\it Top left}: 144 MHz LOFAR map at $10''\times6''$ resolution ($P.A.=81^{\rm o}, \; \sigma \sim 0.15 \; {\rm mJy \; beam^{-1}}$). The western source `L' has been recently reclassified as a radio galaxy. The SE relic is labelled as `J' and a bridge-like structure links it to the source `K'. {\it Top right}: 1.5 GHz JVLA (B, C, D arrays combined) map at $5''\times4''$ resolution ($P.A.=16^{\rm o}, \; \sigma \sim 18 \; {\rm \mu Jy \; beam^{-1}}$) showing the discrete sources embedded in the radio halo that we subtracted. {\it Bottom left}: Zoom of the above LOFAR image on the two extended sources `B' (head-tail radio galaxy) and `C'. {\it Bottom right}: Zoom of the above JVLA image on the radio galaxy `L' showing its FRI morphology.}
	\label{maphigres}
 \end{figure*}

\begin{figure*}
	\centering
	\includegraphics[height=9.5cm,width=8.5cm]{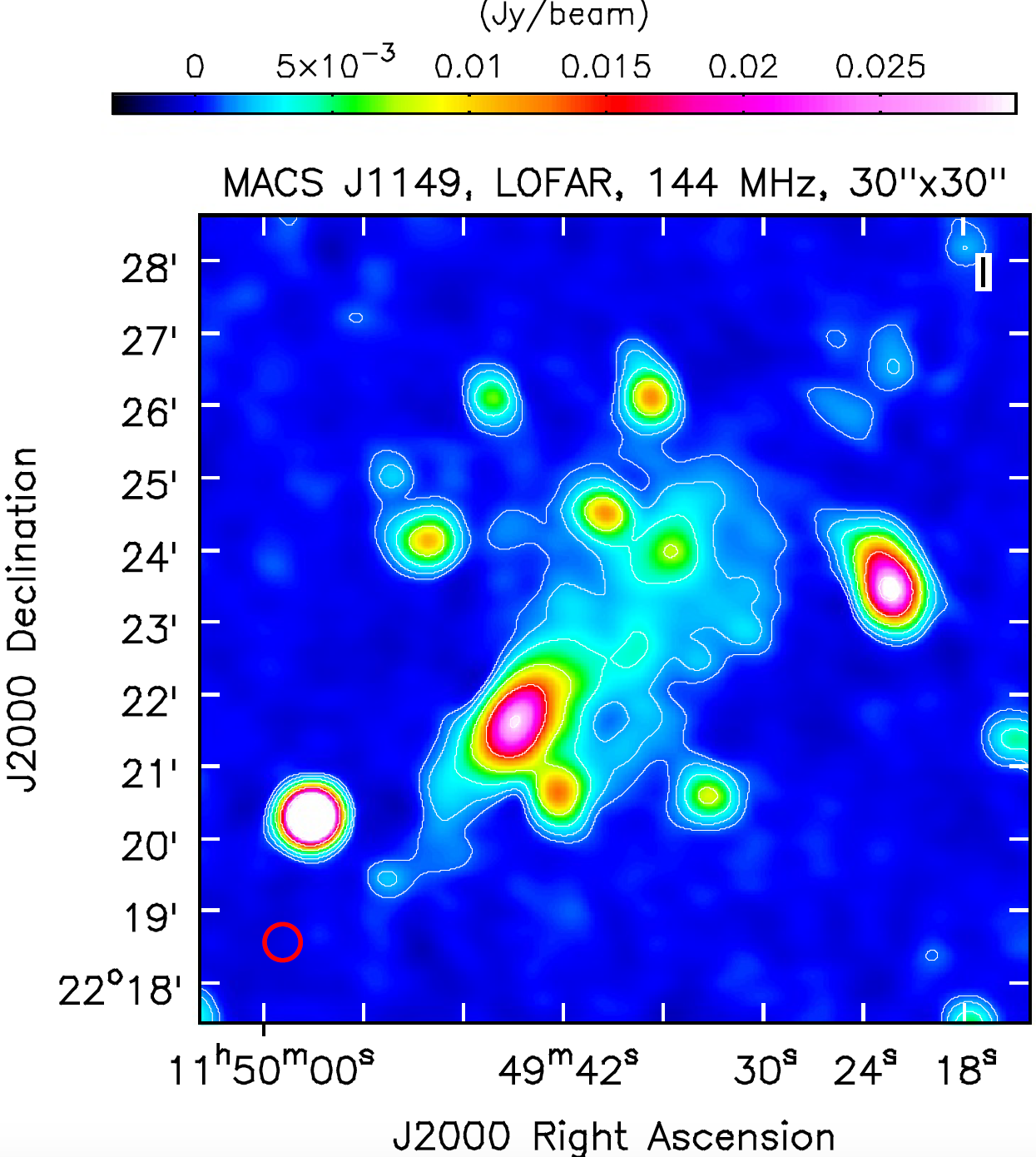} 	
    \includegraphics[height=9.5cm,width=8.5cm]{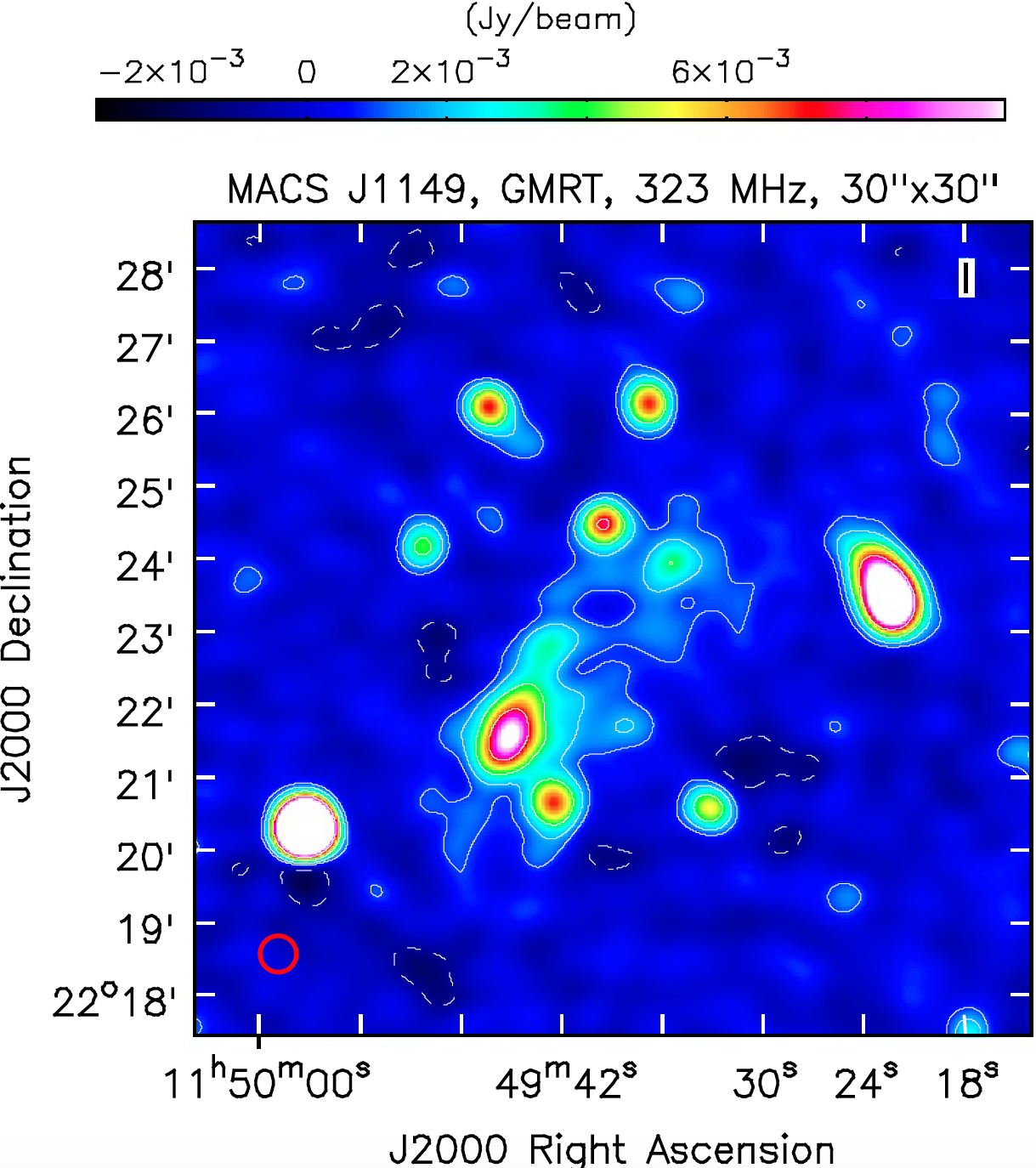}
    \includegraphics[height=9.5cm,width=8.5cm]{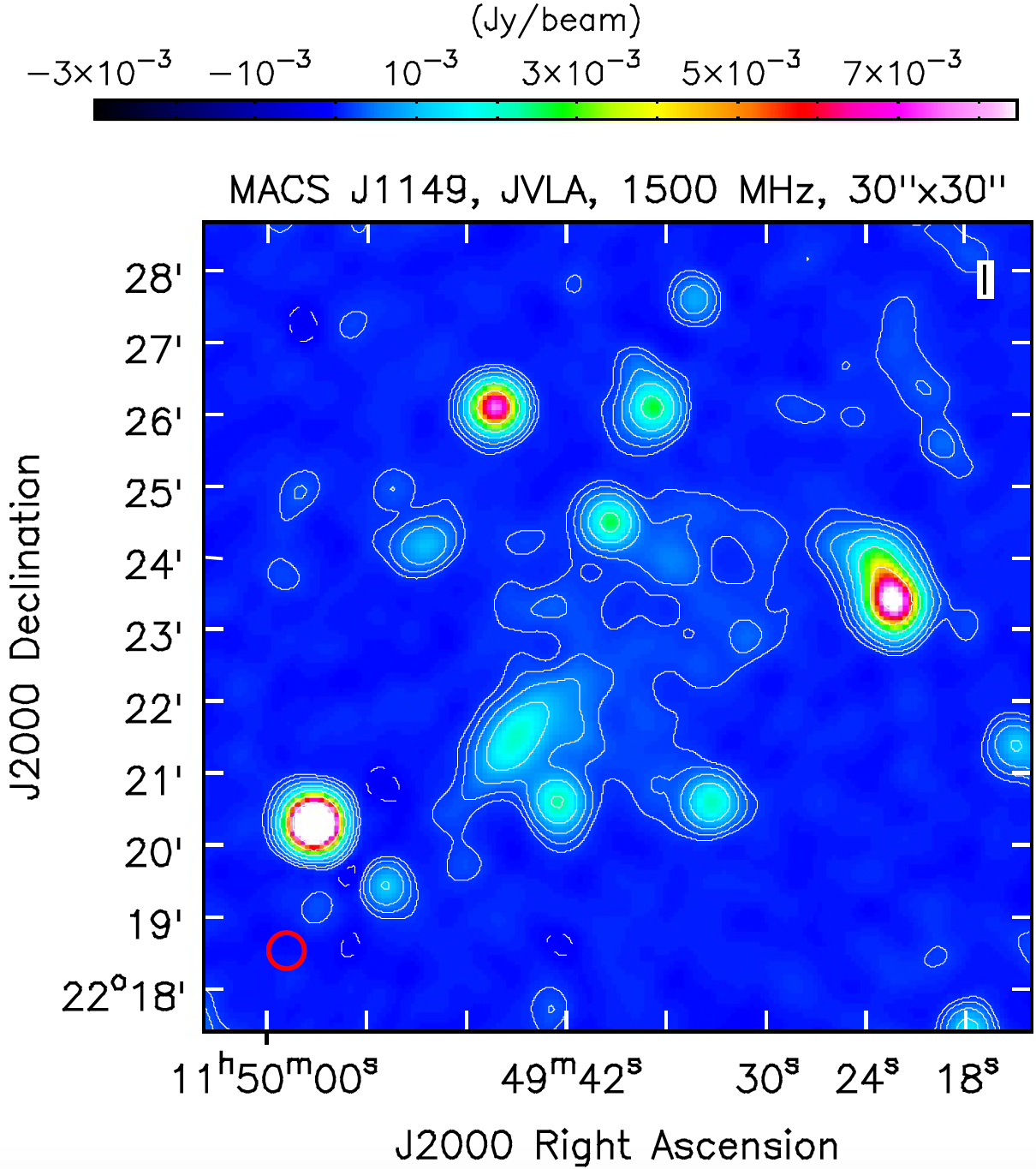} 
    \includegraphics[height=8cm,width=8.5cm]{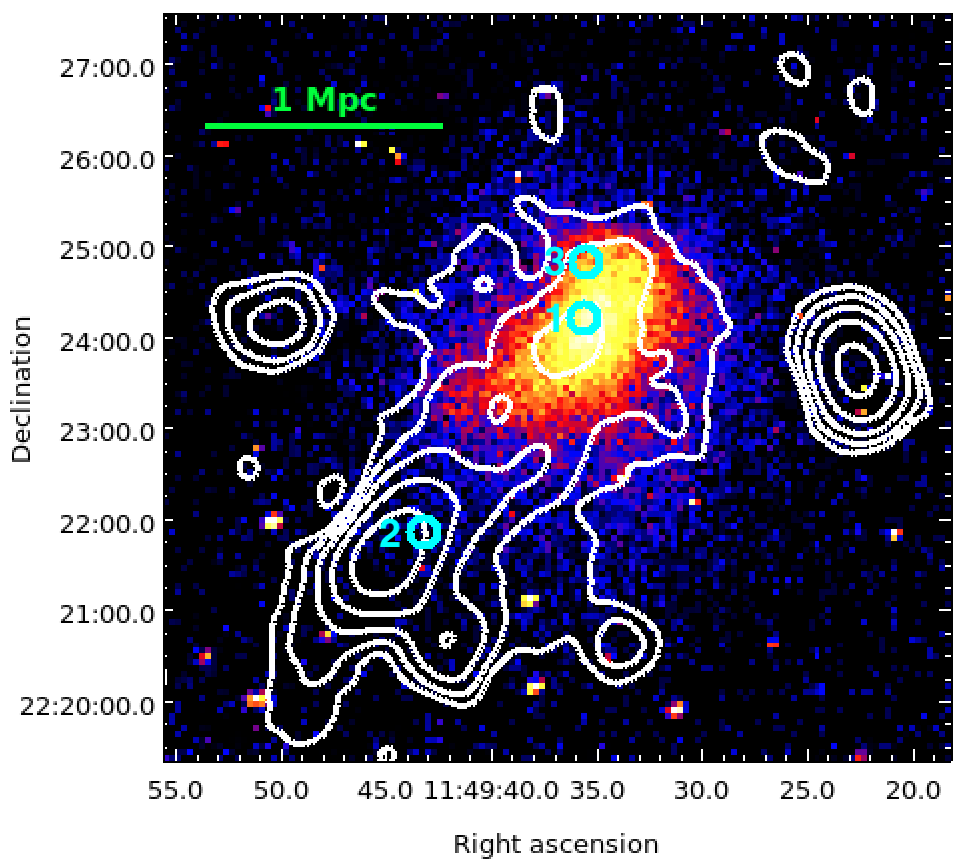}
	\caption{MACS J1149 radio maps at $30''\times30''$ resolution. In all the panels, the contour levels are $[\pm3, \;6, \;12, ...]\times \sigma$. {\it Top left}: 144 MHz LOFAR map ($\sigma \sim 0.30 \; {\rm mJy \; beam^{-1}}$). 
	{\it Top right}: 323 MHz GMRT map ($\sigma \sim 0.30 \; {\rm mJy \; beam^{-1}}$). {\it Bottom left}: 1.5 GHz JVLA (D array) map ($\sigma \sim 45 \; {\rm \mu Jy \; beam^{-1}}$). {\it Bottom right}: 144 MHz LOFAR contours (after the subtraction of the embedded sources) overlaid on the X-ray image of Fig. \ref{mappX2}.}
	\label{mappe}
 \end{figure*}

\begin{figure*}
	\centering
	\includegraphics[height=9cm,width=13cm]{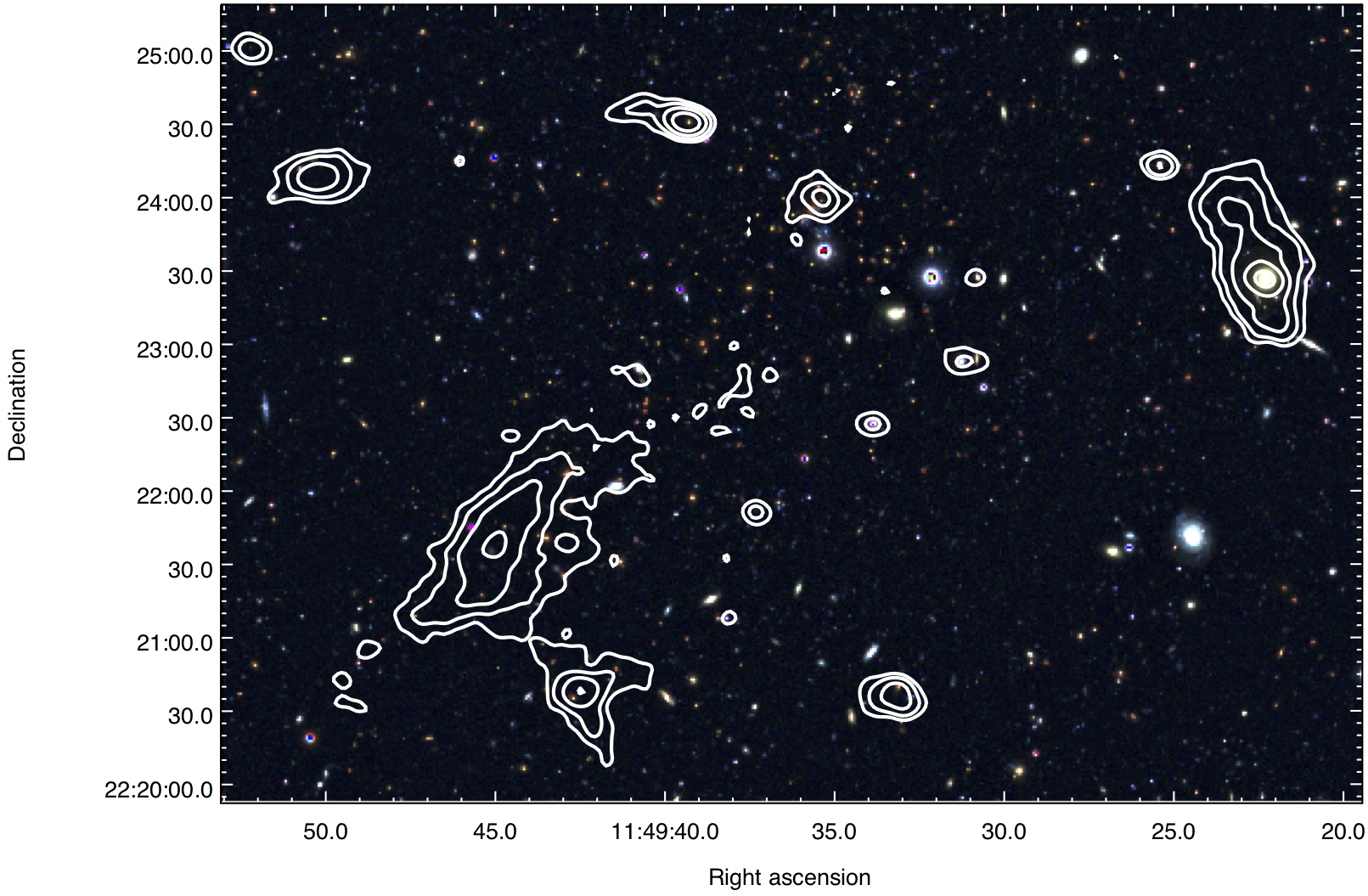}
	\caption{Japanese Virtual Observatory Subaru (RVB filters) image of MACS J1149. The contours of the 144 MHz LOFAR map ($10''\times6''$, $P.A.=81^{\rm o}$)  shown in Fig. \ref{maphigres} are displayed in white.}
	\label{subaru}
 \end{figure*} 

\begin{figure}
	\centering
	\includegraphics[height=5.5cm,width=7cm]{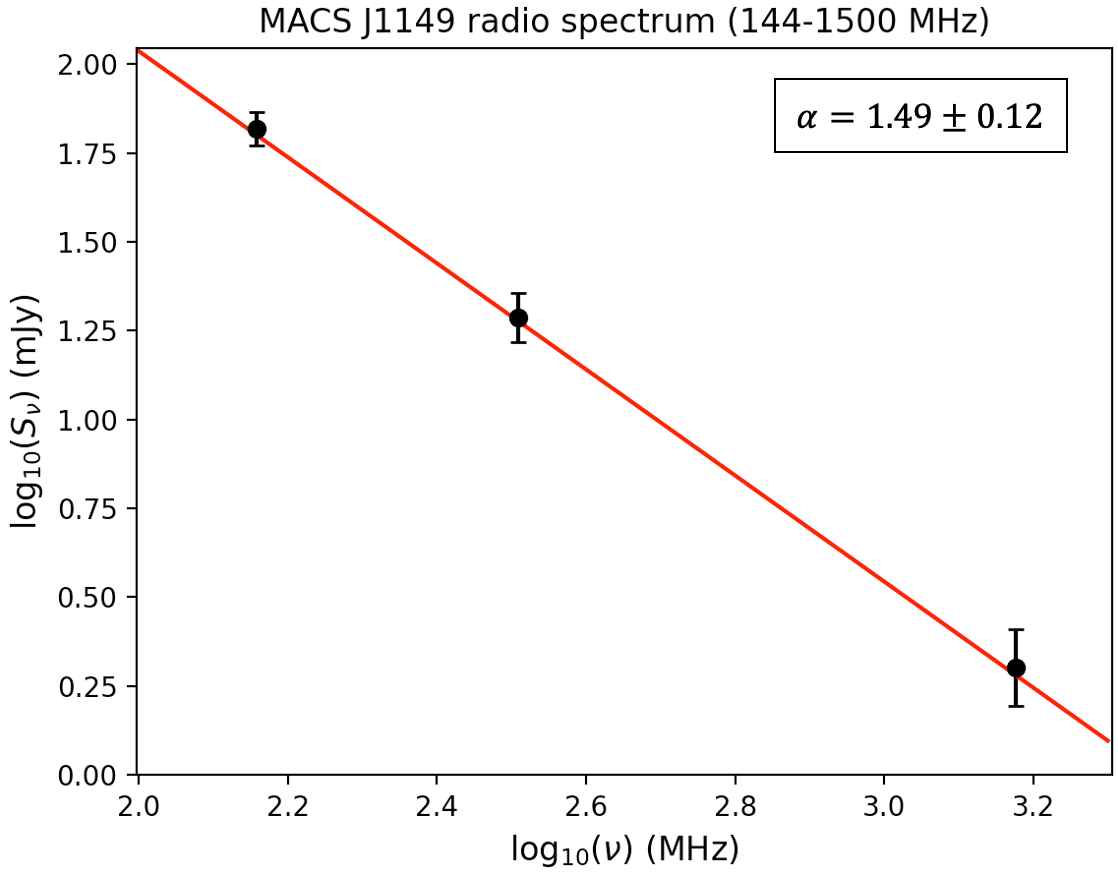}
	\caption{Radio spectrum of the halo between 144 and 1500 MHz (see Table \ref{fluxtabsource}). The solid line indicates the fitted power law. The resulting spectral index is $\alpha_{144}^{1500}=1.49\pm0.12$.}
	\label{spettro}
 \end{figure}

\begin{table}
\centering
	\caption[]{Flux density at 144, 323, and 1500 MHz of the sources shown in Figs. \ref{maphigres} and \ref{mappe}}
	\label{fluxtabsource}   
	\begin{tabular}{cccc}
	\hline
	\noalign{\smallskip}
	Source & $S_{144}$ & $S_{323}$ & $S_{1500}$ 
	\\  
	 & (mJy) & (mJy) & (mJy)  \\  
    \hline
	\noalign{\smallskip}
	Halo & 65.8$\; \pm \;7.1$ & $19.4\; \pm \;3.1$ & $2.0\; \pm \;0.5$  \\
     A &  &  & $0.15\; \pm \;0.01$  \\
     B & $16.6 \; \pm \; 1.7$ & $8.1\; \pm \;0.8$ & $3.1\; \pm \;0.2$  \\
     C & $4.1 \; \pm \; 0.4$ & $3.2\; \pm \;0.3$ & $1.1\; \pm \;0.1$  \\
     D1+D2  & $2.9 \; \pm \; 0.3 $ &  $0.8\; \pm \;0.1$ & $0.38\; \pm \;0.02$   \\
     E &  & $0.7\; \pm \;0.1$ & $0.28\; \pm \;0.01$  \\
     F &  &  & $0.19\; \pm \;0.01$  \\
     G1+G2 &  &  & $0.52\; \pm \;0.03$   \\
     H1+H2 &  &  & $0.31\; \pm \;0.02$  \\
     I & $0.7\; \pm \; 0.1$ &  & $0.16\; \pm \;0.01$  \\
     L & 63.8$\; \pm \;6.4$ & $39.9\; \pm \;4.1$ & $17.1\; \pm \;0.9$  \\
\noalign{\smallskip}
	\hline
	\end{tabular}  
	\begin{tablenotes}
\item    {\small \textbf{Notes}. Sources D1 and D2 are unresolved at 144, 323, and 1500 (D array) MHz. Sources G1, G2, H1, and H2 are unresolved at 1500 (D array) MHz. Missing values mean that the source is undetected at $3\sigma$ level.} 
 \end{tablenotes}	
\end{table}

\begin{figure}
	\centering
	\includegraphics[height=7cm,width=7cm]{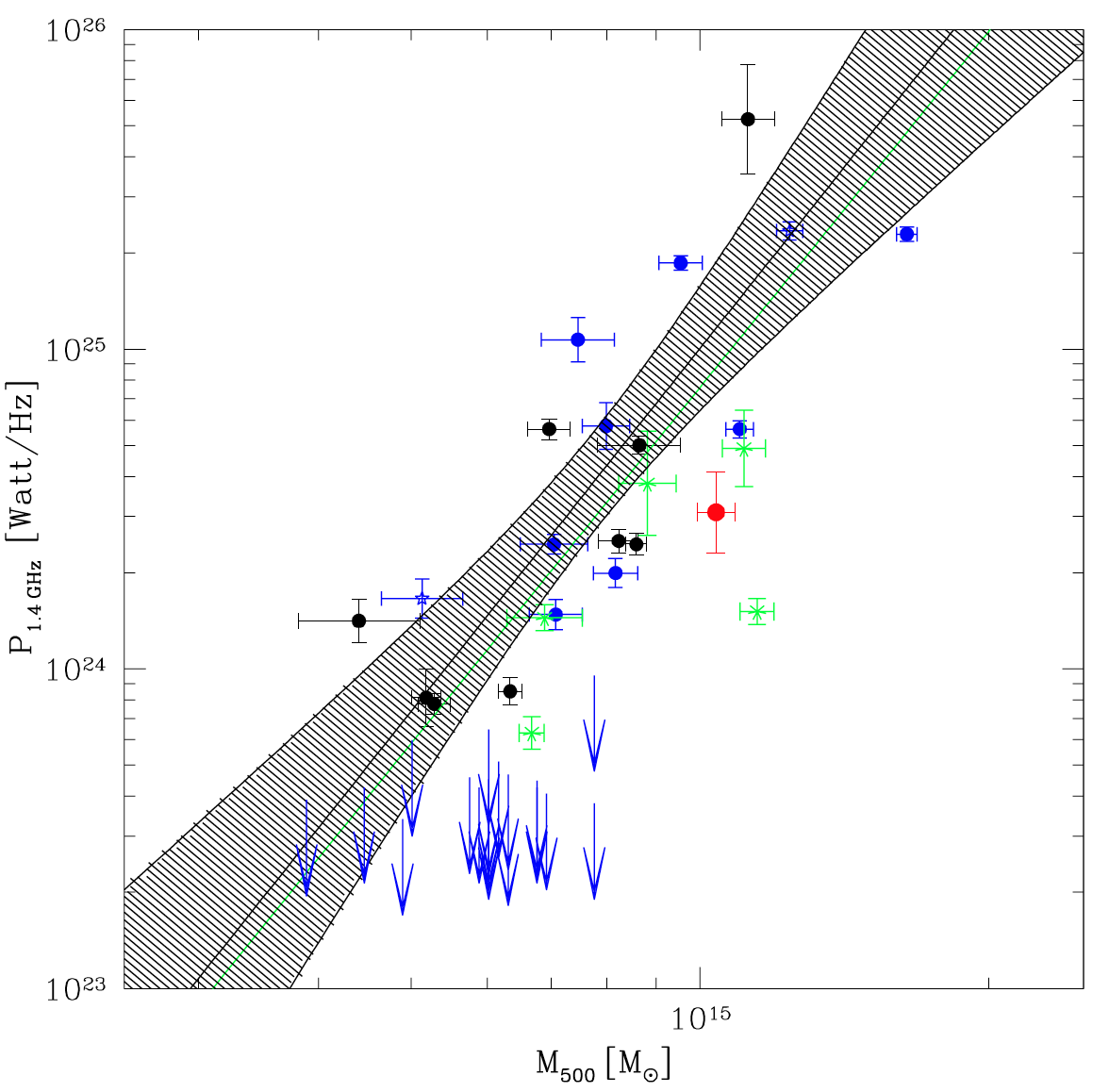}
	\caption{Distribution of galaxy clusters in the $P_{1400}$-$M_{500}$ plane \citep[adapted from][see it for details of the legend]{cassano13}. MACS J1149 (red dot), with $M_{500}=(10.4\pm0.5)\times 10^{14} \; {\rm M_\odot}$ and $P_{1400}=(3.1\pm0.9)\times10^{24} \; {\rm W\; Hz^{-1}}$, is under-luminous with respect to the correlation by a factor of $\sim 3$, and locates among the steep spectrum radio halos (green dots).}
	\label{corr_cassano13}
 \end{figure}

In the left panel of Fig. \ref{maphigres} we present our 144 MHz LOFAR map at $10'' \times 6''$ resolution, while the right panel reports the 1.5 GHz JVLA map at $5'' \times 4''$ resolution, obtained by combining the data in B, C, and D arrays. These high resolution images show the presence of several compact and extended sources. 

The bright elongated sources in the south-east (`J') and west (`L') are the relic and the candidate relic, respectively, reported by \cite{bonafede12}, which are well recovered at both frequencies. As mentioned, the source `L' has been recently reclassified as a radio galaxy \citep{giovannini20}. This is further confirmed by means of our deeper JVLA data: the bottom right panel of Fig. \ref{maphigres} focuses on the source `L' and unambiguously indicates that it is actually a radio galaxy, with a well resolved core and jets, and no hotspots, with a FRI \citep{fanaroff} morphology. We measured a flux density of $S_{144}=63.8 \pm 6.4$ mJy, $S_{323}=39.9 \pm 4.1$ mJy, and $S_{1500}=17.2 \pm 0.9$ mJy, at 144, 323, and 1500 MHz, respectively. We fitted these values with a single power law, obtaining a spectral index $\alpha_{144}^{1500}=0.56\pm0.01$ between 144 and 1500 MHz, consistent with the typical spectra of radio galaxies. We notice that the flux density measured by \cite{giovannini20} at 1.5 GHz is significantly lower (by a factor $\sim 2$); in fact, their maps are shallower (the RMS is $\sim 2$ times higher) than ours and allow them to recover only the brightest inner emission of the source (roughly corresponding to the $12\sigma$ level in the bottom right panel of Fig. \ref{maphigres}).

The halo emission is not clearly detected at the $3\sigma$ level, due to its low surface brightness and the high resolution of the images. There are two extended sources, `B' and `C', highlighted in the bottom left panel. An accurate description of the SE relic `J' is beyond the scope of this work, however we notice that the source `K' at its SW appears appears to be partially embedded in diffuse emission and shows an interesting bridge-like structure linking it with the SE relic. Due to their shallower data, this bridge is not detected by \cite{giovannini20}. It is not clear whether this diffuse emission belongs to the relic or the source itself. As discussed by \cite{giovannini20}, the orientation of `J' is expected to be perpendicular to the merger axis, if it is a relic induced by the merger shock. Therefore, they proposed that `J' might be a radio filament associated with the merger and not a relic.

Fig. \ref{mappe} shows the $30''\times30''$ resolution maps at 144 MHz (top left), 323 MHz (top right), and 1.5 GHz (bottom left), that we used to consistently measure the flux density of the halo. For the first time, we have deep radio maps of the halo at both high and low frequency and a more reliable measure of the spectrum can be performed.
In all these images, the diffuse emission  extends up to $\sim 400'' \; (\sim 2.6 \; {\rm Mpc})$ along the NW-SE direction, even though it is fully recovered only at 144 MHz. The morphology of the halo emission is similar at 144 MHz, 323 MHz, and 1.5 GHz. It has a projected NW-SE length of $\sim 200'' \; (\sim 1.3 \; {\rm Mpc})$ and is interconnected with the relic. In the bottom right panel of Fig. \ref{mappe} we overlay the LOFAR contours (after the subtraction of the sources embedded in the halo) on the X-ray image of Fig. \ref{mappX2}. The radio emission of the halo is co-spatial with the X-ray emission of the ICM, as typically observed in galaxy clusters.

A number of compact sources, labelled in Fig. \ref{maphigres}, and the above mentioned `B' and `C' sources are embedded within the halo. In particular, source `B' at 144 MHz apparently has a head-tail morphology, with the tail barely visible above 323 MHz. Source `C' lies in the centre of the halo and appears extended. In Fig. \ref{subaru} we  overlay the high resolution LOFAR contours on the RVB-filters image of the cluster obtained with Subaru/Suprime-Cam \citep{miyazaki02}. This image suggests a possible association of the source `C' with a cluster member galaxy (coordinates RA$_{\rm J2000} \; 11^h49^{m}35.51^s$, Dec$_{\rm J2000} \;  22^o24'03.76''$).

As discussed in Sect. \ref{wscleansub}, the approaches of subtracting the visibilities and the flux densities of the sources are consistent in the LOFAR and GMRT datasets. However, given the very low flux density of the halo at 1500 MHz, small errors in the subtraction may bias it high, and given the JVLA image properties we have more confidence in the algebraic subtraction. Table \ref{fluxtabsource} summarises the net flux density of the halo and the embedded sources (we also report the source `L'). The obtained values of the halo within the $3\sigma$ LOFAR contours (by avoiding contamination of the regions nearby the relic\footnote{the region where we measured the flux density of the halo roughly corresponds to the one sampled with red boxes in Fig. \ref{radioxcomb}, right panel.}) are $S_{144}=65.8 \pm 7.1$ mJy, $S_{323}=19.4 \pm 3.1$ mJy, and $S_{1500}=2.0 \pm 0.5$ mJy, respectively. The resulting spectrum across 144 and 1500 MHz is reported in Fig. \ref{spettro}. We fitted the flux density values with a single power law, obtaining a spectral index of $\alpha_{144}^{1500}=1.49\pm0.12$. We iterated the measures in different regions of the halo (in particular, within the brightest X-ray counterpart; see Fig. \ref{mappe}, bottom right), consistently finding an average $\alpha \sim1.5$ everywhere and thus confirming that MACS J1149 is a steep spectrum radio halo. We notice that \cite{bonafede12} measured $S_{323}=29 \pm 4$ mJy and $S_{1500}=1.2 \pm 0.5$ mJy ($\alpha_{323}^{1500}=2.1$), whereas \cite{giovannini20} obtained $S_{1500}=0.9 \pm 0.1$ mJy. The inconsistency of these values with our measure at 1.5 GHz derives from the fact that our observations includes D array data that are deeper and more sensitive to the extended emission, thus allowing us to properly recover the emission of the halo at high frequency and providing a more reliable measure of the total flux density; the higher flux density at 323 MHz reported by \cite{bonafede12} is likely due to the different adopted weighting scheme and a larger considered area that might be contaminated by the SE relic.

By using the flux density and the fitted spectral index, we computed the corresponding radio powers as:
\begin{equation}
P_{\nu}=4 \pi D_{\rm L}^{2}S_{\nu}(1+z)^{{\alpha-1}}
\label{radiopower}
\end{equation}
We obtained 
$P_{144}=(9.3\pm1.0)\times10^{25} \; {\rm W\; Hz^{-1}} $, $P_{323}=(2.7\pm0.5)\times10^{25} \; {\rm W\; Hz^{-1}} $, and $P_{1500}=(2.8\pm0.7)\times10^{24} \; {\rm W\; Hz^{-1}} $, at 144, 323, and 1500 MHz, respectively. Radio halos with steep spectrum are found to be radio under-luminous with respect to the correlation reported by \cite{cassano13}, between the 1.4 GHz radio power and mass of the host cluster. By extrapolating from the radio spectrum, the flux density at 1.4 GHz results $S_{1400}=2.2\pm0.6$ mJy, and we calculated a power of $P_{1400}=(3.1\pm0.9)\times10^{24} \; {\rm W\; Hz^{-1}} $ . Indeed, MACS J1149 as well falls below this correlation by a factor of $\sim 3$, as shown with a red dot in Fig. \ref{corr_cassano13}.

\subsection{Thermal-non-thermal correlation}
\label{sectionTnT}

\begin{figure*}
	\centering
    \includegraphics[height=6.2cm,width=7.5cm]{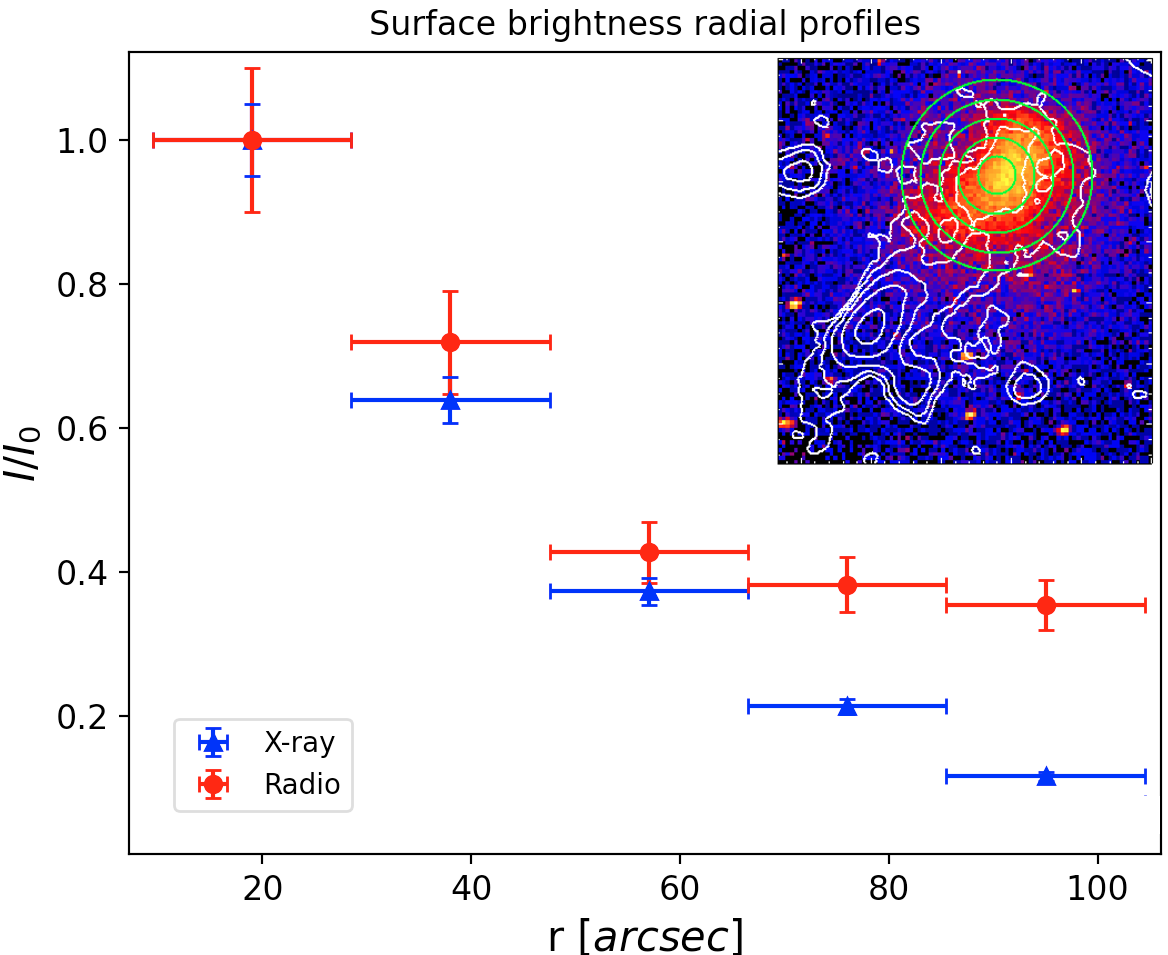}
    \includegraphics[height=6.2cm,width=8.0cm]{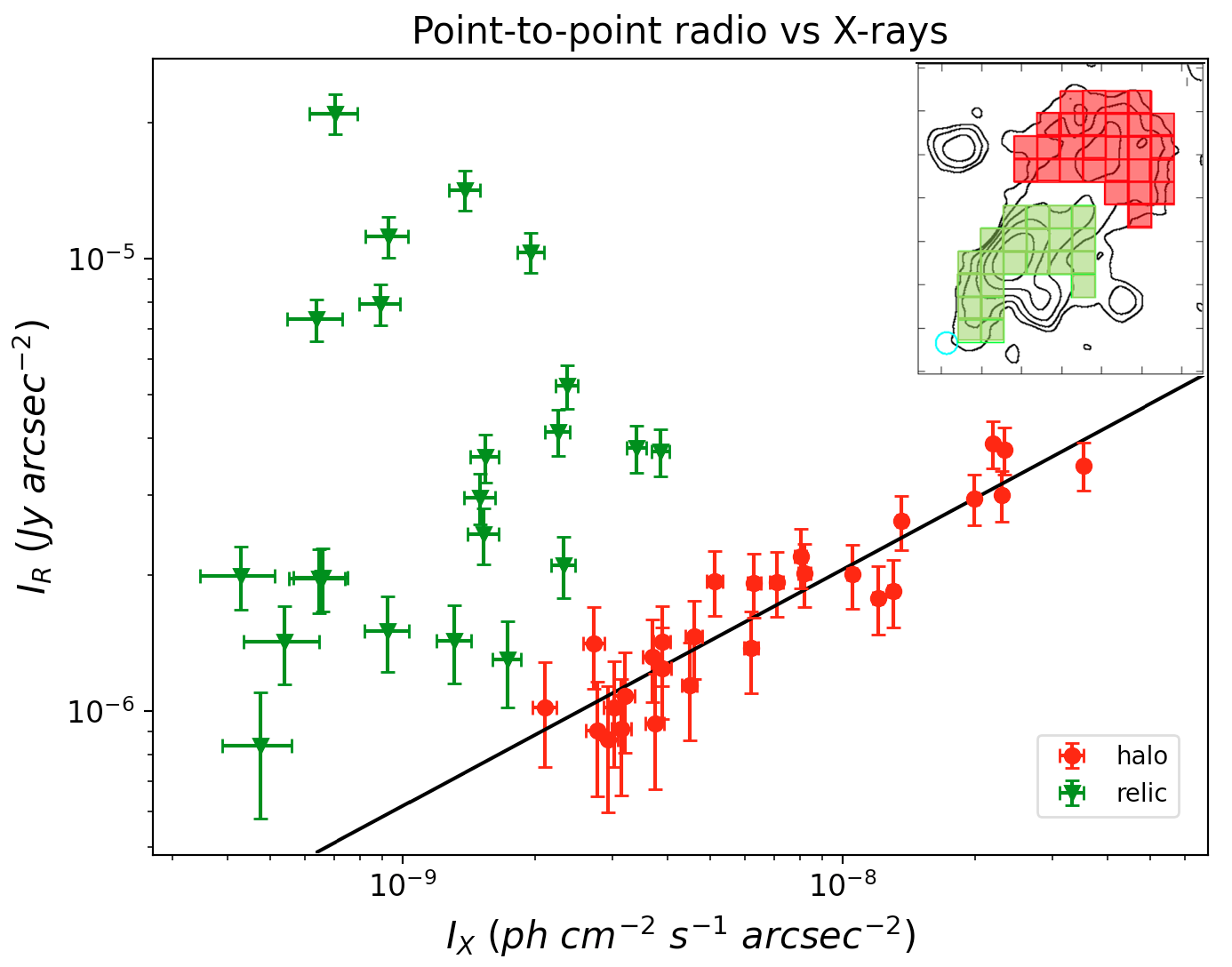}
    \caption{{\it Left}: Radio (red circles) and X-ray (blue triangles) surface brightness radial profiles within the halo region, normalised at their maximum. In the top right corner of the plot we show the adopted source-subtracted LOFAR map at $19''\times15''$ resolution, sampled with concentric annuli of $19''$ width. We remark that $I_{\rm R}$ decreases slower than $I_{\rm X}$, suggesting that they are sub-linearly correlated. {\it Right}: Point-to-point analysis of the $I_{\rm R}$-$I_{\rm X}$ relation. In the top right corner of the plot we report the contours of the $30''\times30''$ source-subtracted LOFAR map, gridded with green and red $30''\times30''$ square boxes for the relic and halo, respectively. The relic points do not show any trend, whereas the halo points present a sub-linear correlation with a slope in the range [0.42-0.60]. The black line represents the best fit obtained with the BCES-orthogonal method at $3\sigma$ threshold.}
	\label{radioxcomb}%
\end{figure*}

\begin{table}
\centering
	\caption[]{Results from the $I_{\rm R}$-$I_{\rm X}$ point-to-point analysis of the halo. We tested {\ttfamily BCES} bisector, {\ttfamily BCES} orthogonal, and {\ttfamily linmix} methods with different radio thresholds. The obtained slopes $k$ are listed in column 3.}
	\label{ptrexfit}   
	\begin{tabular}{ccc}
	\hline
	\noalign{\smallskip}
    Method	& Threshold & $k$  \\  
    \hline
	\noalign{\smallskip}
 &	$2\sigma$ & $0.55\; \pm \;0.04$  \\
 Bisector &	$3\sigma$ & $0.51\; \pm \;0.04$  \\
 &	$4\sigma$ & $0.46\; \pm \;0.04$  \\
 \noalign{\smallskip}
 \hline
  &	$2\sigma$ & $0.56\; \pm \;0.04$  \\
 Orthogonal &	$3\sigma$ & $0.52\; \pm \;0.04$ \\
 &	$4\sigma$ & $0.48\; \pm \;0.04$  \\
 \noalign{\smallskip}
 \hline
  &	$2\sigma$ & $0.53\; \pm \;0.05$  \\
 Linmix &	$3\sigma$ & $0.51\; \pm \;0.05$ \\
 &	$4\sigma$ & $0.48\; \pm \;0.05$  \\
\noalign{\smallskip}
	\hline
	\end{tabular}
	
\end{table}

As typically found in radio halos, the non-thermal emission in MACS J1149 roughly follows the thermal one (Fig. \ref{mappe}, bottom right). The left panel of Fig. \ref{radioxcomb} shows the radio ($I_{\rm R}$), and X-ray ($I_{\rm X}$) surface brightness radial profiles, extracted in circular beam-size annuli (we used the LOFAR image at $19''\times 15''$ resolution) and normalised at their respective maxima. We notice that $I_{\rm R}$ decreases with the radius slower than $I_{\rm X}$. 

A point-to-point correlation of the type of $I_{\rm R}\propto I_{\rm X}^k$ between the radio and X-ray surface brightness is found for a number of radio halos \citep[e.g.][]{feretti01,govoni01,giacintucci05,hoang19,botteon20b,xie20}, with typical sub-linear slopes $k<1$, meaning that $I_{\rm R}$ increases slower than $I_{\rm X}$. We investigated this relation in MACS J1149, by means of the code described in \cite{ignesti20}. The code allows us to sample an extended radio source down to a given threshold in surface brightness, by gridding it with square cells. Both $I_{\rm R}$ and $I_{\rm X}$ are measured within each cell, then the data are fitted with a power law, and the Pearson ($\rho_{\rm p}$) and Spearman ($\rho_{\rm s}$) ranks are computed.

The LOFAR source-subtracted map at resolution $30''\times30''$ was used to grid both the halo and relic with beam-size cells, as represented in the right panel of Fig. \ref{radioxcomb}. Green and red points mark the relic and halo cells, respectively. The plot clearly shows that the halo data follow a tight ($\rho_{\rm p}=0.93$, $\rho_{\rm s}=0.91$) correlation, whereas the data extracted from the region of the relic are uncorrelated. In fact, halos and the ICM are thought to share a similar volume, whereas the volume occupied by relics is much lower (at a first approximation, they can be considered as bidimensional sources), therefore a physical correlation is expected only in the former case. The fact that no correlation seems to exist between the relic and the ICM, allows us to better discriminate the halo region from that of the relic. Nevertheless, the transition region was excluded to avoid any possible contamination to the fit of the correlation slope. 

We explored the correlation strength of the halo through various fitting methods: {\ttfamily BCES} \citep[Bivariate Correlated Errors and intrinsic Scatter;][]{akritasberkshedi96} bisector and orthogonal, and {\ttfamily linmix} \citep{kelly07}. Moreover, we checked for possible biases due to the choice of the radio brightness threshold. The results are listed in Table \ref{ptrexfit}. There is a good agreement among the three methods, but we notice that the fit is sensitive to the adopted threshold, especially in the case of {\ttfamily BCES} bisector and orthogonal. Even though the values of $k$ are consistent, lower thresholds give steeper indices, as recently discussed in \cite{botteon20b}. Anyway, we find that the $I_{\rm R}$-$I_{\rm X}$ correlation is strongly sub-linear in MACS J1149, with $k$ in the range [0.42-0.60].

\section{Discussion}

In this Section, we discuss the formation process of the halo and its connection with the cluster dynamics. 

\begin{figure}
	\centering
	
	\includegraphics[height=7cm,width=7cm]{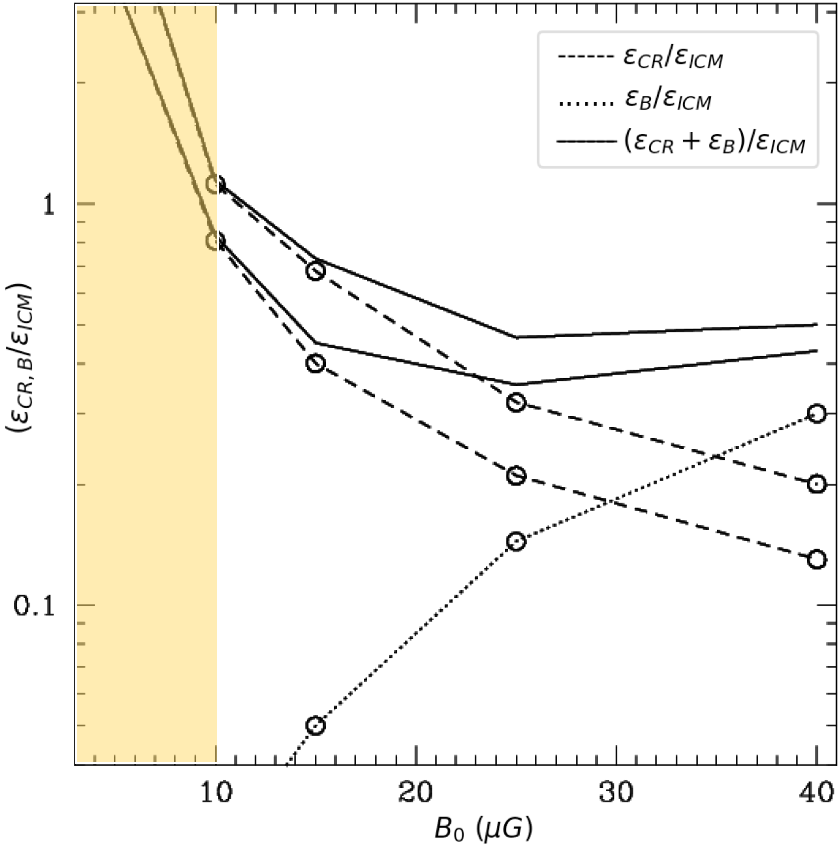} 
	\caption{Non-thermal to thermal energy budget as a function of the central magnetic field in a pure hadronic model. The dashed, dotted, and solid lines are the CRe to ICM energy ratio, the magnetic to ICM energy ratio, and the total non-thermal to ICM energy ratio, respectively, required to match the observed radio emission up to radii of 2.2$r_{\rm c}$ (lower dashed and solid lines) and 2.5$r_{\rm c}$ (upper dashed and solid lines). The non-thermal energy budget for typical values of $B_{0}\lesssim10 \; \mu$G in galaxy clusters is indicated by the yellow area.
	A pure hadronic model would require unrealistically high values of $\left( \varepsilon_{\rm CR}+\varepsilon_{\rm B}\right)/\varepsilon_{\rm ICM}$ and/or $B_{0}$.}
	\label{adronici}%
\end{figure}

\subsection{Constraining the hadronic contribution}
\label{sectionadronici}

In general, hadronic scenarios are disfavoured by the observed connection between radio halos and the cluster dynamics \citep[e.g][]{brunettijones14,vanweeren19}. In a few cases, the spectral properties of the halos \citep[e.g.][]{brunetti08,wilber18} and the limits to the gamma-ray emission of the hosting clusters \citep{brunetti12,brunetti17,zandanel&ando14} provide constraints that are sufficiently stringent to allow pure hadronic models to be excluded. The radio halo in MACS J1149 combines a high redshift with a steep radio spectrum (Fig. \ref{spettro}) and a flat radial profile (Fig. \ref{radioxcomb}, left) which overall challenge an hadronic origin. In the following, we show that the properties of MACS J1149 allow us to test hadronic models without using complementary constraints from gamma-rays.

We aim at estimating the non-thermal energy in the cluster required to reproduce the observed radio emission in a pure hadronic scenario. The stationary spectrum of electrons and positrons produced by hadronic collisions in the ICM is \citep{dolag&ensslin00}:
\begin{equation}
N_{\rm e}^\pm(p,t)=\frac{1}{\sum_{\rm rad, i}^{} \left| \frac{{\rm d}p}{{\rm d}t} \right|}\int_{p}^{} Q_{\rm e}^\pm(p,t)p{\rm d}t 
\label{spettrosecondari}
\end{equation}
where $p$ is the momentum of the particles, $\left| \frac{{\rm d}p}{{\rm d}t} \right|_{\rm rad, i}$ are the radiative and Coulomb losses, and $Q_{\rm e}^\pm(p,t)$ is the injection rate of electrons and positrons defined as \citep[][and references therein]{brunettijones14}:
\begin{equation} 
\begin{split}
 Q_{\rm e}^\pm(p,t) = & \; 2.1712\frac{m_\pi^2 n_{\rm th}c^2}{m_\pi^2-m_\mu^2}\int_{E_{\rm min}}^{}\int_{p_*}^{} \frac{{\rm d}E_\pi {\rm d}p}{E_\pi \Bar{\beta}_\mu} \beta_{\rm p} N_{\rm p}(p,t) \; \times \\
                      & \times \; \frac{{\rm d}\sigma^{\pm,0}}{{\rm d}E}(E_\pi,E_{\rm p}) F_{\rm e}(E_{\rm e}, E_\pi)
\end{split}
\label{spettrosecondari2}
\end{equation}
where the pedices '$\pi$', '$\mu$', and 'p' refer to pions, muons, and protons, respectively, $\Bar{\beta}_\mu=\sqrt{1-\left( \frac{m_\mu}{\Bar{E}_\mu} \right)^2}$, $\Bar{E}_\mu=\frac{E_\pi}{0.5428}\left[ 1-\left(  \frac{m_\mu}{m_\pi} \right)^2 \right]$, and $\frac{{\rm d}\sigma^{\pm,0}}{{\rm d}E}(E_\pi,E_{\rm p})$ 
is the differential inclusive cross-section for the production of $\pi^+$, $\pi^-$, and $\pi^0$   \citep[we assume the cross-section as in][]{brunetti17}, and  $F_{\rm e}(E_{\rm e}, E_\pi)$ is defined as in \cite{brunetti&blasi05} (Eqs. 36 and 37). 

We derived the distribution of the thermal protons, $n_{\rm th}(r)$, from the X-ray surface brightness profile $I_{\rm X}(r)$ (index $\beta=0.66$, core radius $r_{\rm c}=288$ kpc, and central proton density number $n_{\rm th,0}=3.6\times10^{-3} \; {\rm cm^{-3}}$; see Eqs. \ref{betamodeleq} and \ref{betamodelSB}). We also assumed that the magnetic field scales with the thermal proton density as:
\begin{equation}
B(r)=B_0 \left[ \frac{n_{\rm th}(r)}{n_{\rm th,0}}\right]^\eta 
\label{magneticfield}
\end{equation}
where we fixed $\eta=0.5$, being the value which gives the equilibrium between the thermal and magnetic energy densities \citep[e.g.][]{bonafede10}. 

The secondary CRe injected in the ICM magnetic field produce synchrotron radio emission with an emissivity given by \citep{dolag&ensslin00}:
\begin{equation} 
\begin{split}
 j_{\rm R}(\nu,r) & = \sqrt{3} \frac{e^3}{m_{\rm e}c^2}\int_{0}^{\frac{\pi}{2}} \sin^2{\theta}{\rm d}\theta \int_{}^{} N_{\rm e}^\pm(p)F\left(\frac{\nu}{\nu_{\rm c}}\right){\rm d}p \propto \\
& \propto N_{\rm p}(p,r)n_{\rm th}(r)\frac{B(r)^{1+\alpha}}{B(r)^2+B_{\rm CMB}^2} \nu^{-\alpha}
\end{split}
\label{spettrosecondari3}
\end{equation}
where $\theta$ is the pitch angle between $B$ and $p$, $F\left(\frac{\nu}{\nu_{\rm c}}\right)$ is the synchrotron kernel \citep{rybicki&lightman79}, and $B_{\rm CMB}=3.25(1+z)^2 \; \mu $G is the CMB equivalent magnetic field. Under the assumption of a spherical geometry, the integral of $j_{\rm R}(\nu,r)$ along the line of sight provides the surface brightness profile at a projected distance $b$:
\begin{equation}
I_{\rm R}(b)=\frac{1}{4\pi(1+z)^4}\int_{b}^{+\infty} \frac{2rj_{\rm R}(r)}{\sqrt{r^2-b^2}}{\rm d}r
\label{spettrosecondari4}
\end{equation}

By assuming a value of the central field $B_0$ and taking into account the effects of the $k$-correction, we matched the radio profile from Eq. \ref{spettrosecondari4} with the observed one in order to derive the momentum distribution of CRp in the form \citep{brunettijones14}:
\begin{equation}
N_{\rm p}(p,r)=K_{\rm p}(r)p^{-s}
\label{spettrosecondari5}
\end{equation}
where $K_{\rm p}(r)$ is the normalisation and $s=2\alpha$ (with $\alpha=1.5$, inferred from the radio spectrum of the halo). Thus, the energy budget of the CRp is obtained as: 
\begin{equation}
\varepsilon_{\rm p}=\int_{}^{} {\rm d}^3r \int_{p_{\rm min}}^{} EN_{\rm p}(p,r){\rm d}p
\label{spettrosecondari6}
\end{equation}
where we consider a minimum momentum of the CRp spectrum of $p_{\rm min}=0.5m_{\rm p}c$ (corresponding to a kinetic energy of $E_{\rm min}\sim110$ MeV). 

The ratio of CRp and thermal energy budget required to match the observed radio emission assuming a pure hadronic model is shown in Fig. \ref{adronici} as a function of $B_0$. We also show the ratio of the thermal ($\varepsilon_{\rm ICM}$) and non-thermal ($\varepsilon_{\rm CR}+\varepsilon_{\rm B}$) energy budget, integrated within maximum radii of $2.2r_{\rm c}$ and $2.5r_{\rm c}$, where the halo is clearly visible (see Fig. \ref{mappe}). For typical values of $B_0<10 \; {\rm \mu G}$ in galaxy clusters \citep[e.g.][]{bonafede10}, an untenable $\varepsilon_{\rm CR}/\varepsilon_{\rm ICM}>1$ is required. The CRe energy budget declines with increasing $B_0$, however the minimum $\left( \varepsilon_{\rm CR}+\varepsilon_{\rm B}\right)/\varepsilon_{\rm ICM}$ is reached for unusually high values of the magnetic field $B_0\sim25 \; {\rm \mu G}$. Moreover, the CRe energy budget is still extremely large ($\sim 40-50 \%$): for a comparison, the Fermi satellite constrains it to few percent in Coma and other nearby galaxy clusters \citep{ackermann14,brunetti17}. 

Under the assumption of a spherical symmetry of the ICM and CRp distribution, the most important parameter is the minimum energy of CRp. If we conservatively assumed $E_{\rm min}=300$ MeV, i.e. the minimum threshold for CRp-p collisions, $\varepsilon_{\rm CR}$ would just result $30\%$ smaller than in Fig. \ref{adronici}; on the other hand, extending the CRp spectrum at lower energies down to $E_{\rm min}=20$ keV and $E_{\rm min}=1$ MeV would boost up $\varepsilon_{\rm CR}$ by a factor of 3.4 and 2.3, respectively. For these reasons, we can rule out an hadronic origin of the halo; by considering $\varepsilon_{\rm CR}/\varepsilon_{\rm ICM}$ to be few percent, secondary electrons would contribute to $<10\%$ of the halo emission. Nevertheless, we cannot exclude that hadronic collisions still play a role by injecting secondary particles that are eventually re-accelerated by turbulence \citep{brunetti&lazarian11,pinzke17}.

\subsection{Merger scenario}
Lensing and spectroscopy studies concluded that MACS J1149 consists of at least three main sub-clusters \citep{smith09,golovich16}, whose locations are indicated by open circles in Fig. \ref{mappX2}. Sub-clusters 1 and 2 are the most massive ones, with virial masses of $M_{\rm 1} \sim M_{\rm 2} \sim 10^{15} \; {\rm M_{\odot}}$, whereas sub-cluster 3 is one order of magnitude less massive $M_{\rm 3} \sim 10^{14} \; {\rm M_{\odot}}$. 

The ICM has a clearly elongated structure along the NW-SE axis and although sub-clusters 1 and 2 have similar masses, only the first one is associated with the main X-ray concentration. According to the scenario proposed by \cite{golovich16}, a major merger between sub-clusters 1 and 2 occurred $\sim 1$ Gyr ago along the NW-SE direction in the plane of the sky and sub-cluster 2 was stripped of its gas. The two X-ray tails of Fig. \ref{ggm} in the south could be related to this event. Besides the major merger between sub-clusters 1 and 2, a minor merger between sub-clusters 1 and 3 has begun $\sim 100$ Myr ago, likely along the line of sight \citep{golovich19}. Given the high mass ratio, this interaction does not significantly affect the ICM and the major merger dynamics.

In proximity of the N-NE candidate cold front identified by \cite{ogrean16}, we find instead that there may be two distinct X-ray surface brightness discontinuities in NE and NW possibly being cold fronts. However, the case of Abell 3667 \citep{owers09,ichinohe17} suggests a simpler scenario in which we are observing a single cold front, where the coolest gas was pushed towards the NW, and is surrounded by the hotter gas. Under this assumption, our cold front can likely be explained with the major merger described above, and its origin could be associated with the cool core of a sub-cluster. However, because of the low count statistic in the upstream regions, at present we cannot disentangle the nature of the discontinuities.

The major merger likely induced the formation of the relic and the halo through shocks and turbulence. Finding that the candidate misaligned W relic actually is a FRI radio galaxy is in line with this scenario. The steepening of the spectral index of the SE relic towards the centre of the cluster \citep{bonafede12} would also be consistent with this model, however its orientation is peculiar, as it is parallel to the merger axis and not perpendicular, as observed for relic sources. The interaction of a merger shock with remnants or clouds filled by pre-existing relativistic electrons may explain this feature. On the other hand, \cite{giovannini20} suggested that the SE source might be a radio filament associated with the halo. Nevertheless, with our data we do not exclude the possibility that the SE source might be a radio galaxy; this will be the subject of a future paper.

Steep spectrum radio halos are predicted as a result of less efficient turbulent acceleration or stronger energy losses \citep[e.g.][]{brunetti08,cassano10A}. The former case may result from a less energetic sequence of merger events \citep[e.g.][]{cassano06} or may mark the final evolution phase of energetic mergers where a significant fraction of the turbulence has already been dissipated \citep{donnert13}. In both cases, halos are predicted to be under-luminous. The old major merger in MACS J1149 may be in line with the advanced  evolution phase scenario, however it is difficult to discriminate between the two possibilities.

\section{Summary and conclusions}
In this work, we constrained the spectrum of the radio halo in the high redshift ($z=0.544)$ MACS J1149 galaxy cluster, by means of multifrequency LOFAR, GMRT, and JVLA radio data at 144, 323, and 1500 MHz, and found it to be steep. In addition, we used Chandra X-ray data to compare the thermal and non-thermal emission of the target.

MACS J1149 is a merging system consisting of at least three sub-clusters. The X-ray data analysis showed that the ICM is characterised by a remarkably high temperature ($kT=10.5\pm 0.3$ keV) and luminosity ($L^{500}_\mathrm{[0.1-2.4 \; keV]}=(1.35\pm0.01)\times10^{45} \; \mathrm{erg\; s^{-1}}$), and is elongated along the NW-SE axis, where the two most massive sub-clusters lie. Despite its large mass, no X-ray clumps are associated with the SE sub-cluster. 

We found two surface brightness discontinuities, that we interpret as the evidence of a temperature gradient within a single cold front, where the coolest gas was pushed towards NW. In this case, the cold front may be explained as the cool core remnant of a sub-cluster involved in the major merger occurred $\sim 1$ Gyr ago. A higher SNR in the upstream regions is required to better constrain the nature of the discontinuities.   

We produced radio images at different frequency and resolution, showing that the halo is co-spatial with the ICM. The measured flux densities provide a spectral index $\alpha_{144}^{1500}= 1.49\pm 0.12$ between 144 and 1500 MHz. The computed radio powers of the halo are $P_{144}=(9.3\pm1.0)\times10^{25} \; {\rm W\; Hz^{-1} }$, $P_{323}=(2.7\pm0.4)\times10^{25} \; {\rm W\; Hz^{-1} }$, and $P_{1500}=(2.8\pm0.7)\times10^{24} \; {\rm W\; Hz^{-1} }$. As typical steep spectrum halos, MACS J1149 is under-luminous (by a factor of $\sim 3$) with respect to the correlation found by \cite{cassano13} between the 1.4 GHz radio power versus the mass of the cluster.   

A point-to-point comparison of the radio and X-ray surface brightness showed a tight sub-linear correlation $I_{\rm R} \propto I_{\rm X}^k$, which we investigated through various fitting methods and radio thresholds, consistently finding a slope $k$ in the range [0.42-0.60].  

Finally, thanks to the high redshift of the cluster, the steep spectrum of the halo, and its flat spatial distribution, we were able to test hadronic models without constraints from gamma-ray observations. Our analysis demonstrated that a significant contribution to the halo from pure hadronic collisions is ruled out. In line with previous studies, the old ($\sim 1$ Gyr ago) major merger between the two most massive sub-clusters likely induced turbulence in the ICM, which accelerated particles and formed the halo. The steep spectrum of the halo may results both from low turbulence in combination with high radiative (inverse Compton) losses or an advanced evolution phase. 

In addition, we confirm the recent reclassification proposed by \cite{giovannini20} of the W source as a radio galaxy, and in particular we point out its FRI morphology. The peculiar orientation of the SE relic may indicate a different nature of this source and requires a more detailed analysis, that is beyond the aim of this work.

\begin{acknowledgements}
We thank the referee for his comments and suggestions, that have improved the presentation of the paper. 
KR acknowledges financial support from the ERC Starting Grant 'MAGCOW', no. 714196. GB, FG, and RC acknowledge support from INAF mainstream project 'Galaxy Clusters Science with LOFAR', 1.05.01.86.05. A. Botteon acknowledges support from the VIDI research program with project number 639.042.729, which is financed by the Netherlands Organisation for Scientific Research (NWO). AI acknowledges the Italian PRIN-Miur 2017 (PI A. Cimatti). A. Bonafede acknowledges support from the ERC-Stg DRANOEL n. 714245 and from the Italian MIUR grant FARE 'SMS'. RJvW and GDG acknowledge support from the ERC Starting Grant ClusterWeb, no. 804208. VC acknowledges support from the Alexander von Humboldt Foundation. MB acknowledges support from the Deutsche Forschungsgemeinschaft under Germany's Excellence Strategy - EXC 2121 "Quantum Universe" - 390833306.
LOFAR \citep{vanhaarlem13} is the Low Frequency Array designed and constructed by ASTRON. It has observing, data processing, and data storage facilities in several countries, which are owned by various parties (each with their own funding sources), and that are collectively operated by the ILT foundation under a joint scientific policy. The ILT resources have benefited from the following recent major funding sources: CNRS-INSU, Observatoire de Paris and Universit\'e d’Orl\'eans, France; BMBF, MIWF- NRW, MPG, Germany; Science Foundation Ireland (SFI), Department of Business, Enterprise and Innovation (DBEI), Ireland; NWO, The Netherlands; The Science and Technology Facilities Council, UK; Ministry of Science and Higher Education, Poland; The Istituto Nazionale di Astrofisica (INAF), Italy. This research made use of the Dutch national e-infrastructure with support of the SURF Cooperative (e-infra 180169) and the LOFAR e-infra group. The J\"ulich LOFAR Long Term Archive and the German LOFAR network are both coordinated and operated by the J\"ulich Supercomputing Centre (JSC), and computing resources on the supercomputer JUWELS at JSC were provided by the Gauss Centre for Supercomputing e.V. (grant CHTB00) through the John von Neumann Institute for Computing (NIC). This research made use of the University of Hertfordshire high-performance computing facility and the LOFAR-UK computing facility located at the University of Hertfordshire and supported by STFC [ST/P000096/1], and of the Italian LOFAR IT computing infrastructure supported and operated by INAF, and by the Physics Department of Turin University (under an agreement with Consorzio Interuniversitario per la Fisica Spaziale) at the C3S Supercomputing Centre, Italy. 
The National Radio Astronomy Observatory is a facility of the National Science Foundation operated under cooperative agreement by Associated Universities, Inc. We thank the staff of the GMRT that made these observations possible. GMRT is run by the National Centre for Radio Astrophysics of the Tata Institute of Fundamental Research. The scientific results reported in this article are based on observations made by the Chandra X-ray Observatory data obtained from the Chandra Data Archive. This research has made use of data and/or software provided by the High Energy Astrophysics Science Archive Research Center (HEASARC), which is a service of the Astrophysics Science Division at NASA/GSFC. Based in part on data collected at Subaru Telescope, which is operated by the National Astronomical Observatory of Japan. Part of the data are retrieved from the JVO portal (http://jvo.nao.ac.jp/portal) operated by the NAOJ.

\end{acknowledgements}

\bibliographystyle{aa}
\bibliography{bibliografia}


\end{document}